\documentclass[journal=jacsat,manuscript=article]{achemso}

\usepackage{chemformula} 
\usepackage[T1]{fontenc} 
\usepackage{float}  %
\usepackage{siunitx}

\usepackage[normalem]{ulem}

\author{Zhihao Zhou}

\author{Khushboo Kumari}
\affiliation{Department of Electrical and Computer Engineering, University of Washington, Seattle, Washington 98195, USA}

\author{Ningzhi Xie}
\affiliation{Department of Electrical and Computer Engineering, University of Washington, Seattle, Washington 98195, USA}

\author{Shane Colburn}
\affiliation{Department of Electrical and Computer Engineering, University of Washington, Seattle, Washington 98195, USA}
\alsoaffiliation{Office of the Director of National Intelligence (ODNI)}

\author{Chetan Poudel}
\affiliation{Department of Chemistry, University of Washington, Seattle, Washington 98195, USA}

\author{Praneeth Chakravarthula}
\affiliation{Department of Computer Science, University of North Carolina at Chapel Hill, Chapel Hill, North Carolina 27599, USA}

\author{Karl F. B\"ohringer}
\affiliation{Department of Electrical and Computer Engineering, University of Washington, Seattle, Washington 98195, USA}
\alsoaffiliation{Department of Bioengineering, University of Washington, Seattle, Washington 98195, USA}
\alsoaffiliation{Institute for Nano-Engineered Systems, University of Washington, Seattle, Washington 98195, USA}

\author{Arka Majumdar}
\affiliation{Department of Electrical and Computer Engineering, University of Washington, Seattle, Washington 98195, USA}
\alsoaffiliation{Department of Physics, University of Washington, Seattle, Washington 98195, USA}

\author{Johannes E. Fr\"och}
\email{jfroech@uw.edu}
\affiliation{Department of Electrical and Computer Engineering, University of Washington, Seattle, Washington 98195, USA}

\title{Multifunctional Imaging and Sensing Miniscope with Meta-optics}
\title{Meta-optical Miniscope for Multifunctional Imaging}

\keywords{miniscope, imaging, meta-optics, metasurface}

\begin{document}

\begin{tocentry}

\includegraphics[width=8cm,height=4cm]{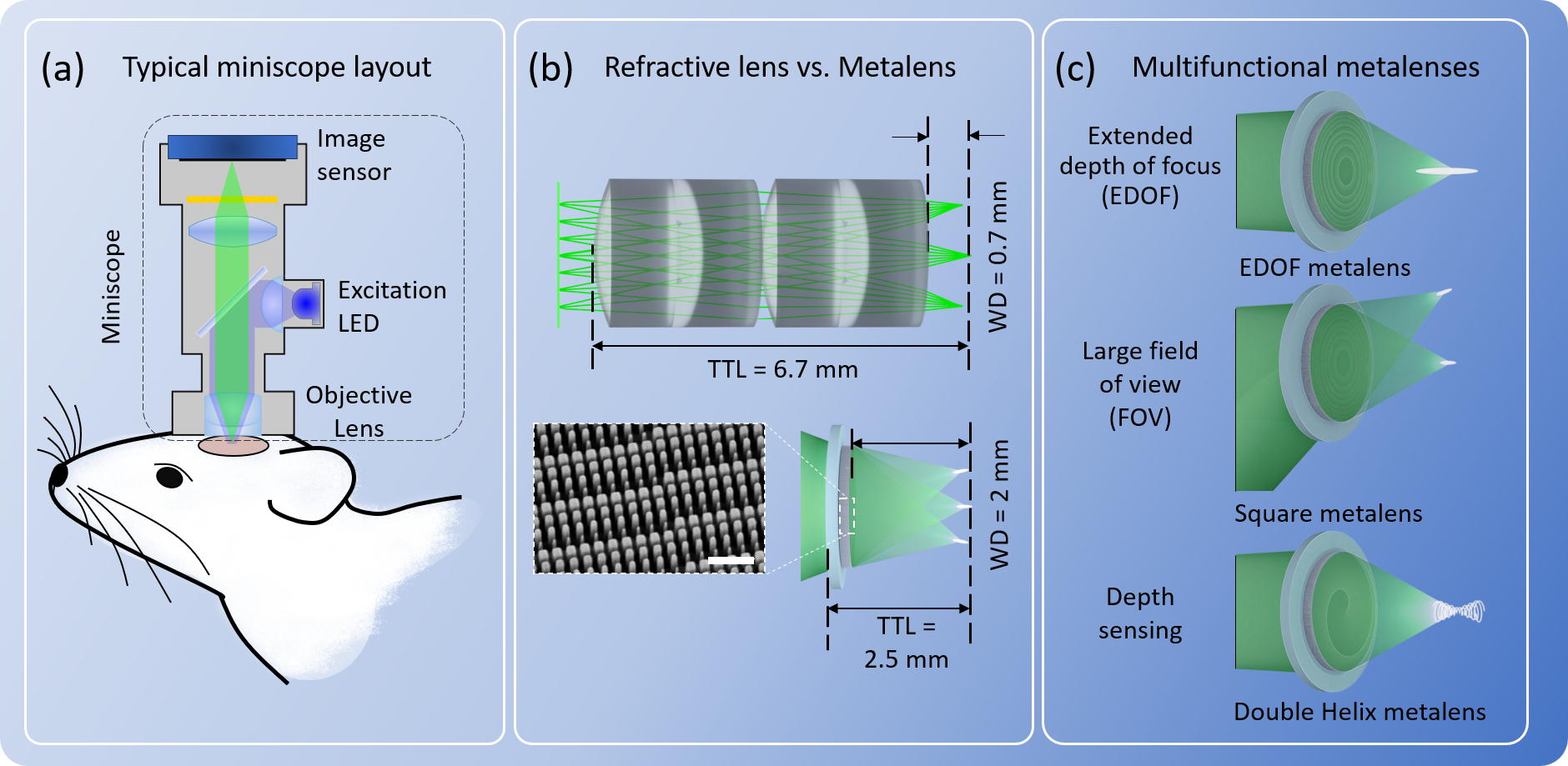}
\end{tocentry}

\begin{abstract}
Miniaturized microscopes (miniscopes) have opened a new frontier in animal behavior studies, enabling real-time imaging of neuron activity, as animals remain mostly unconstrained with the ability to perform a multitude of tasks. Canonical designs utilize either Gradient-Index (GRIN) lenses or a set of refractive lenses as the objective module for excitation and collection of fluorescence. However, GRIN lenses are often limited in optical performance due to aberrations, whereas refractive lenses provide superior optical performance but suffer from bulky and complex designs. Meta-optics, which are sub-wavelength diffractive optical elements, have shown potential to replace traditional solutions while also combining multiple functionalities and significantly reducing the overall footprint and weight of a system. In this work, we leverage the diverse functionalities that meta-optics can enable through their design, including large field of view (FOV), extended depth of focus (EDOF), and depth sensitivity, which expands the miniscope tool-box. These meta-optics are seamlessly integrated by replacing the traditional refractive lens assembly. Performance evaluations highlight the enhanced imaging capabilities, aligned with pre-designed functionalities. Furthermore, the reduced thickness of the meta-optics can shorten the total track length of the objective module from \SI{6.7}{\milli\metre} to \SI{2.5}{\milli\metre}. These improvements have the potential to facilitate the development of more compact and multifunctional miniscopes.

\end{abstract}

\section{Introduction} 

\begin{figure}[htbp]
    \centering
    \includegraphics[width=\linewidth]{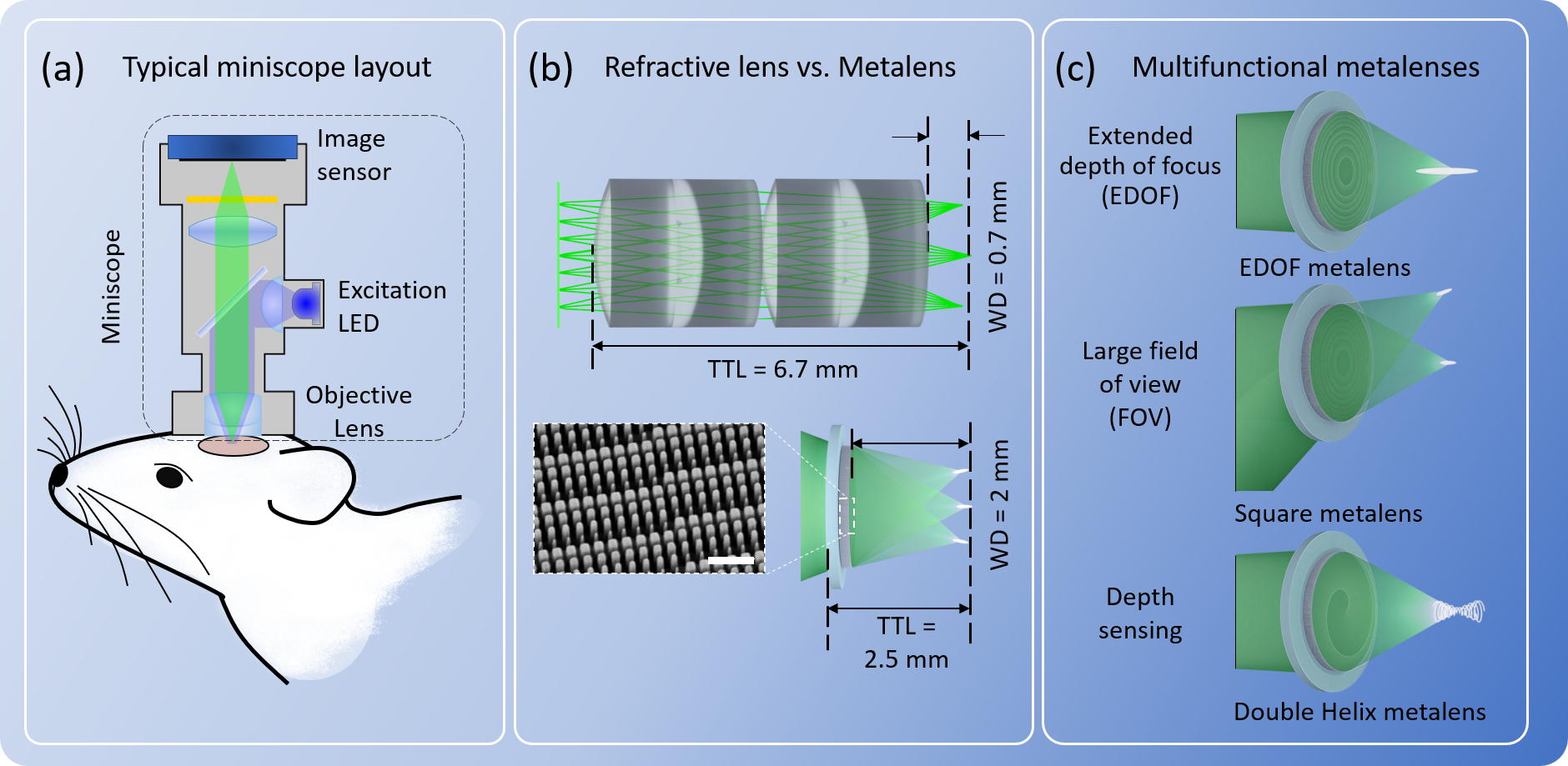}
    \caption{Expansion of the miniscope tool-box with smaller optics and multiple functionalities. (a) Schematic illustration of a miniscope, as currently deployed. (b) Comparison of a refractive assembly and a metalens as integrated with the miniscope, showing the difference in total track length (TTL) and working distance (WD) of the objective lens in a traditional miniscope with those of a metalens, highlighting the reduction in system footprint. Inset scale bar corresponds to \SI{1}{\micro\metre}. (c) Each metalens serves as the objective lens and is engineered to provide distinct imaging functionalities: the extended depth of focus (EDOF) metalens extends the depth of focus (DOF), the square metalens enables a large field of view (FOV), and the double-helix (DH) metalens supports depth sensing. The animal illustration in (a) is included only to depict the typical placement of a miniscope and does not indicate in-vivo imaging performed in this study.}
    \label{fig1}
\end{figure}

Fluorescence microscopes\cite{lichtman2005fluorescence} are essential tools in biological studies. Initially limited to ex-situ imaging of stained tissue, fluorescence microscopy have evolved to enable in-vivo imaging\cite{rao2007fluorescence, barretto2009vivo}, such as monitoring neural activity. However, this typically requires large setups, which contain numerous optical components, including excitation source, dichroic mirror, objective lens, tube lens, filters, and sensors. These complex and bulky systems are unsuitable for head-mounted imaging in animals, limiting their use in freely behaving studies. To overcome this limitation, miniaturized fluorescence microscopes (miniscopes)\cite{ghosh2011miniaturized} have been developed. By integrating essential optical components into a compact, millimeter-scale form factor, miniscopes enable head-mounted operation and real-time in-vivo imaging of neural activity\cite{stamatakis2021miniature, chen2022advances}. Their low power consumption further supports extended observation sessions. Current imaging system designs typically use a pair of objective and tube lenses, which constitute a critical component of the miniscope to simultaneously provide high imaging quality and compact size \cite{aharoni2019circuit}.  

Early miniscopes adopted a gradient index (GRIN) lens as the objective due to its compactness and ease of integration\cite{ghosh2011miniaturized}. Subsequent designs, such as Guo et al.'s triplet achromatic system\cite{guo2023miniscope}, improved field of view (FOV) and resolution. Nevertheless, these approaches remain reliant on refractive optics, which are typically only available as off-the-shelv solutions, thus limiting design flexibility and demand careful control over alignment. Despite these advancements, the development of miniscopes with enhanced functionality or streamlined designs remains largely unexplored. Only in recent years have various research groups substantially expanded the functionality of the miniscope. For example, Liberti et al.\cite{liberti2017open} introduced an open-source, lightweight miniscope with optional wireless data transmission, alleviating tethering constraints during behavioral experiments. Miniscopes with 3D imaging capability over small volumes have also been realized by placing a microlens array at the focal plane of the tube lens, following the principle of light-field microscopy \cite{prevedel2014simultaneous, skocek2018high}, or by replacing the tube lens with a random phase mask \cite{yanny2020miniscope3d}. Subsequent works introduced diffractive optical elements to extend the depth of focus (DOF) \cite{xue2020single, greene2023pupil}. Wide-FOV miniscopes have been demonstrated through optical design optimization\cite{scherrer2023optical, guo2023miniscope, zhao2025minixl} or computational reconstruction\cite{yang2024wide}. Towards extreme miniaturization, the FlatScope\cite{adams2017single,adams2022vivo} offers a much smaller footprint and enables 3D imaging through a lensless design, but its resolution is limited for dense samples and real-time reconstruction is not yet feasible.

Meta-optics, a class of ultra-flat subwavelength diffractive elements, is emerging as a promising platform for integration and diverse optical functionalities\cite{kuznetsov2024roadmap}. Their thickness is fundamentally limited only by the height of the nanoscatterer itself ($\sim$~0.1 - 1~\si{\micro\metre}) and the substrate containing these ($\sim$~0.1 – 0.5 mm). They offer precise control over phase, amplitude, and polarization\cite{hsiao2017fundamentals}, enabling optical functions beyond the reach of conventional refractive optics. These include tunable focal length \cite{wei2021varifocal,colburn2019simultaneous}, customized vectorial focal curves \cite{wang2021metalens}, wide FOV imaging \cite{lin2024wide,wirth2025wide} and augmented reality visors\cite{lee2018metasurface,gopakumar2024full,tian2025achromatic}. Owing to their compactness and compatibility with semiconductor fabrication, meta-optics can be integrated with refractive lenses\cite{zang2024inverse}, microdisplays\cite{fan2024integral}, and spectral filters\cite{mcclung2020snapshot} to enhance imaging, enable 3D displays, and support snapshot hyperspectral imaging. Importantly, recent advances in metalenses for functional imaging systems further highlight the versatility of meta-optics \cite{kim2025rapid,kuang2024palm,shen2023monocular,badloe2023bright}, offering distinct advantages in reducing system volume while enabling additional functionalities. Moreover, meta-optics allow for co-design with computational backends, which can leverage synergistic effects of both doamins and augment data capture and reconstruction, beyond capabilities given by refractive systems only\cite{froch2025computational}. 

In this work, we introduce the "metascope"—a modified miniscope integrating various classes of meta-optics to expand its imaging modalities. As shown in Figure~\ref{fig1}a, we replace the refractive objective of the UCLA Miniscope V4\cite{daniel_aharoni_2023_7844004} with a planar, single-layer metalens. The advantage of the ultrathin meta-optics profile is highlighted in Figure~\ref{fig1}b, showing how this system achieves an extended working distance (WD, from 0.7~mm to 2~mm) and a significantly reduced total track length (TTL, from 6.7~mm to 2.5~mm), enabling a more compact footprint. The various design types used in this work, illustrated in Figure~\ref{fig1}c are tailored for specific functionalities: a meta-optic with extended depth of focus (EDOF) to image across a large volume\cite{cao2023optical}, a square metalens to allow for imaging with large FOV\cite{martins2020metalenses}, and a meta-optic with a double-helix (DH) point spread function (PSF) for depth sensing\cite{colburn2020metasurface,hao2024single}. We first validate each design through PSF analysis. We then evaluate the imaging performance of the miniscope and the metascope equipped with the hyperbolic, square, and EDOF metalenses using a resolution target as well as several biological samples, including mouse kidney sections, multilayer fibrous tissue, and 1.9~\si{\micro\metre} fluorescent beads. Finally, we use the 1.9~\si{\micro\metre} fluorescent beads to assess the depth-sensing capability of the metascope equipped with the DH metalens.

The results confirm that the metascope enables EDOF, wide-FOV imaging, and depth-resolved sensing. In contrast, the traditional refractive-based miniscope achieves high-quality imaging, which however is limited to the focal plane, with significant degradation at other imaging depths or towards larger FOV. Notably, customizing or upgrading the system is straightforward, requiring only the replacement of the objective lens, which essentially allows researcher to explore additional functionalities, beyond those presented in this paper. The main contribution of this work lies in demonstrating a practical and versatile platform that implements metalenses in an application suited for compact imaging. Our study proves that while neither refractive nor metasurface-based optics alone can provide a universal solution for all imaging requirements, the integration of metalenses offers strong functional complementarity. Importantly, we provide a clear quantitative comparison between different imaging modalities, and their limitations. Moreover, we demonstrate the excellent compatibility of our metasurface design with commercially available miniscope systems, achieving significant performance enhancement and functionality extension with only minimal modifications to the existing setup.

\section{Results and discussion}

\subsection{Design and fabrication}

\begin{figure}
    \centering
    \includegraphics[width=\linewidth]{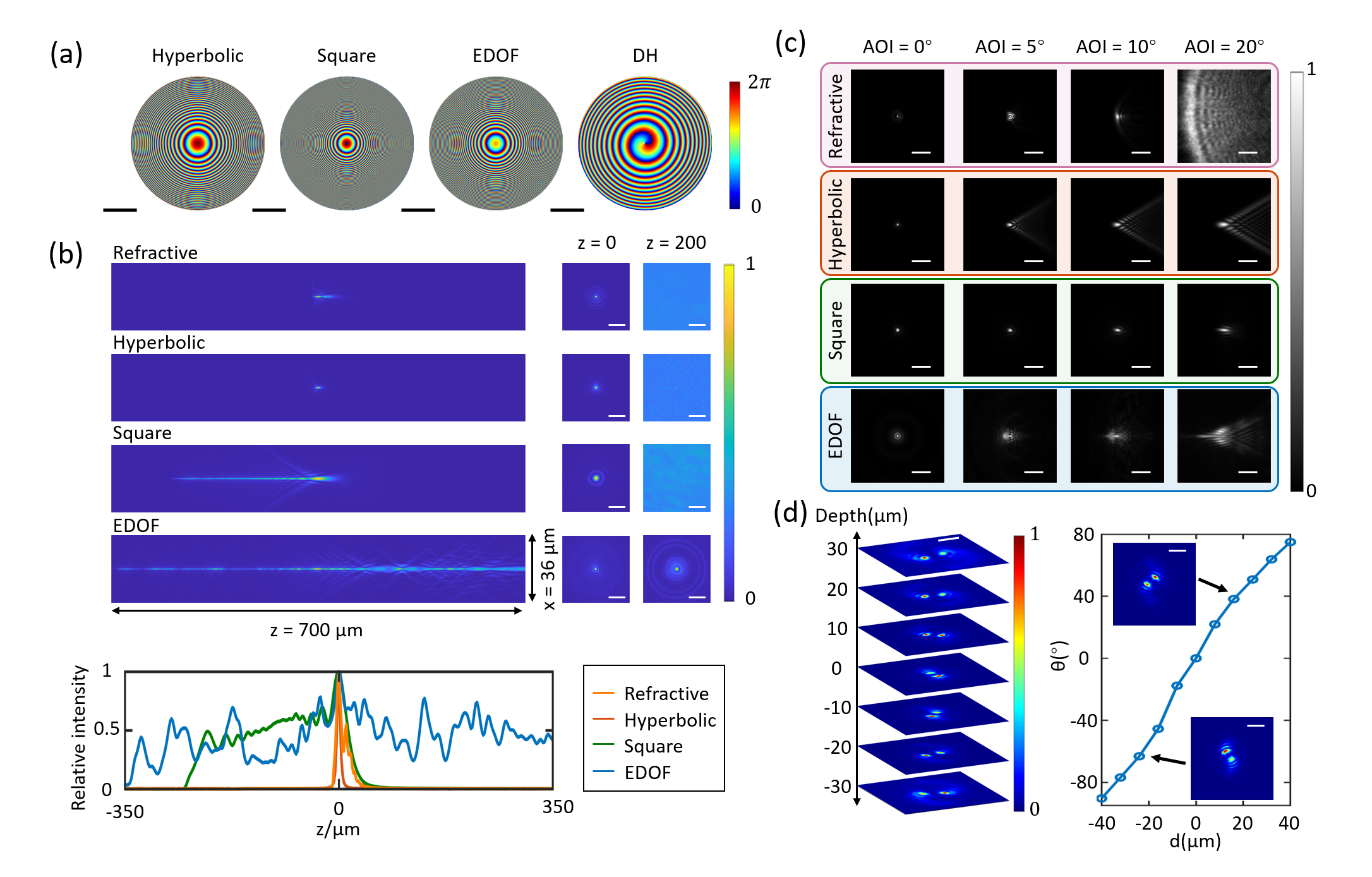}
    \caption{Phase profiles and point spread function (PSF) of the refractive lens and meta-optics. (a) The wrapped phase profiles of the central regions of the hyperbolic, square, EDOF, and DH designs. (b) Cross-sectional intensity distributions ($x$–$z$ and $x$–$y$ planes) and relative normalized intensity curves for the refractive, hyperbolic, square, and EDOF designs at different axial depths. (c) PSFs under varying angle of incidence (AOI, 0° to 20°) for the refractive, hyperbolic, square, and EDOF designs. (d) PSF of the DH meta-optic at different depths and the relationship between the rotation angle and depth. Scale bars: (a): 50~\si{\micro\metre}; (b): 5~\si{\micro\metre}; (c), (d): 10~\si{\micro\metre}.}
    \label{fig2}
\end{figure}

The current imaging system of the miniscope \cite{daniel_aharoni_2023_7844004} (Miniscope-v4) (Figure~\ref{fig1}a) includes an excitation light-emitting diode (LED), a dichroic mirror, an objective lens, a tube lens, and an electrowetting lens for focus adjustment. This version differs from previous iterations primarily in its implementation of the objective lens, which consists of two pairs of achromatic lenses (\#45-089, Edmund Optics) instead of a GRIN lens. This improves image quality with reduced aberrations, specifically towards off-axis rays, which are common to GRIN lenses\cite{lee2011adaptive,easum2016analytical}. At the same time, their functionality and FOV is severely limited. Herein, we show how these limitations can be overcome by considering different designs that provide extended functionalities, as summarized in Figure~\ref{fig1}c. Specifically, we engineer the phase profile, using hyperbolic and square phase, as well as metalenses designed for EDOF and DH PSF. The metalens incorporating a hyperbolic phase profile effectively eliminates spherical aberration, making it a well-established benchmark for evaluating focusing performance \cite{khorasaninejad2016metalenses}. Its phase profile is given by:

\begin{equation}
\varphi(r,\lambda) = -\frac{2\pi}{\lambda} \left( \sqrt{f^2 + r^2} - f \right)
\end{equation}

In contrast, the metalens with a square phase profile enhances the FOV by breaking rotational symmetry and replacing it with translational symmetry \cite{pu2017nanoapertures,martins2020metalenses}. The corresponding phase profile is expressed as:

\begin{equation}
\varphi(r,\lambda) = -\frac{\pi r^2}{\lambda f}
\end{equation}

Here, $r$ represents the radial position, while $\lambda$ and $f$ correspond to the working wavelength and focal length, respectively. In the design, we assume an aperture of 2 mm in diameter with a focal length of 2 mm, for a wavelength of 530 nm, corresponding to the typically collected fluorescence wavelength in miniscopes, under excitation with the integrated blue LED. Small-f-number lenses are employed to highlight the functional differences between metalens designs, as the divergence in their phase profiles becomes more pronounced at large numerical apertures.\cite{liang2019high}. For practical applications, the design parameters of the metalens, including its diameter and focal length, play a crucial role in determining the overall system performance. We have therefore added a more detailed discussion on how these parameters influence the imaging performance, as presented in the Supporting Information. We note that changing these parameters is straightforward, enabling adaptation of a large variety of custom focal lengths/diameters. In comparison, prototyping refractive lenses or GRIN lenses can be significantly more challenging, and the choice is typically limited to an already available selection of lenses. To achieve the EDOF capability, we employed an inverse design approach \cite{bayati2020inverse,froch2025beating}, where the phase profile is optimized such that the targeted sum of logarithmic intensities along the defined DOF is maximized. Specifically, we optimized the PSF, to extend continuously over a 700 \si{\micro\metre} depth range, rather than concentrating solely on the primary focal plane. The detailed optimization procedure, including the field propagation to different focal planes and the definition of the loss function, can be found in the Supporting Information. The DH metalens produces a PSF with two lobes rotating along a helix, whereas the relative angle of rotation can be used to measure the source depth\cite{yang2023realizing,hao2024single}. We optimized the intensity of the two rotating lobes at different angles across varying planes by fine-tuning the coefficients of the clouds of a Laguerre-Gauss mode. The phase profile was then combined with the standard hyperbolic phase profile. Further details on the optimization process can be found in the relevant literature. \cite{jin2019metasurface,colburn2020metasurface,fröch2022dual} 

Figure~\ref{fig2}a illustrates the phase profiles for the different metalenses. To translate these phase profiles to a physical structure, we first created a lookup table, which allows us to correlate a specific phase value to a specific feature size of nanopillar. We simulated the phase response of silicon nitride pillars on a fused silica substrate with a thickness of 800 nm and a center-to-center spacing of 350 nm using rigorous coupled-wave analysis (RCWA). The optimized phase profiles were subsequently converted into feature maps, using this lookup table. These designs were then fabricated using a nanofabrication approach, which is described in detail in the Materials and Methods section and shown in Figure S1a of the Supporting Information. An alumina hard mask was deposited via evaporation after patterning the silicon nitride layer by electron-beam lithography (EBL), followed by inductively coupled plasma reactive ion etching. The optical and scanning electron microscope (SEM) images of the fabricated EDOF metalens (Figure~S1b,c) demonstrate good fabrication quality. 

\subsection{Optical performance characterization}

We first characterized all metalenses on a home-built translatable microscope setup, consisting of a Nikon objective lens (100× magnification, NA = 0.90), a tube lens (Thorlabs TTL200MP), and a camera (ASI2600MM Pro), mounted together on a programmable translation stage (ESP301-3G). Light from a supercontinuum laser (SuperK FIANIUM, NKT Photonics) was passed through a tunable filter, providing narrowband illumination at 530 nm, and was then fiber coupled. The fiber output was mounted and collimated on a rotation stage (Thorlabs RP03) to produce a collimated beam with a controllable angle of incidence (AOI). A schematic diagram of the experimental setup is provided in Figure S2. This setup allowed us to capture the PSF along a large DOF and FOV. Intensity cross-sections along the optical axis and the lateral plane, as well as the relative normalized intensity distribution across a depth of -350 \si{\micro\metre} to 350 \si{\micro\metre} under normal incidence illumination, are shown in Figure~\ref{fig2}b. Here, the DOF is defined as the axial range over which the peak intensity remains above 50$\%$ of its maximum value. As designed, the refractive lens and hyperbolic metalens exhibit a short DOF, while the square metalens provides a slightly larger DOF \cite{pu2017nanoapertures} of about 142 \si{\micro\metre}. In comparison, the EDOF metalens maintains an extended intensity distribution over an exceptionally large depth range of about 266 \si{\micro\metre}. We also present the respective PSF cross-sections at sections at $z = 0$~\si{\micro\metre} and $z = 200$~\si{\micro\metre}. At the design focal length, all metalenses exhibit a distinct focal spot. However, the square metalens and EDOF metalens demonstrate prominent side lobes due to spherical aberration. At $z = 200$ \si{\micro\metre}, the EDOF metalens maintains a clear focal spot, whereas no distinct focal spot is observed for the other lenses.

To compare the performance of different lenses with respect to the FOV, we fixed the detection plane at the focal plane and varied the AOI from 0$^\circ$ to 20$^\circ$. The results are shown in Figure~\ref{fig2}c. The refractive lens and the hyperbolic metalens exhibit minimal aberrations at 0$^\circ$. As the AOI increases, the hyperbolic metalens develops pronounced coma aberration, although the focal spot remains identifiable. The refractive lens maintains a reasonably good PSF up to 10$^\circ$, but its performance rapidly degrades beyond this angle. In contrast, the EDOF metalens exhibits strong field aberrations over all AOIs, whereas the square metalens shows reduced aberrations relative to the other designs, although its PSF becomes slightly elongated along the illumination direction. More detailed comparisons of the DOF and FOV, using metrics such as the Strehl ratio, FWHM, and modulation transfer function (MTF), are provided in the Supporting Information. 

The focusing properties and the relative angular change in axis of the two lobes of the DH metalens is also discussed in Figure~\ref{fig2}d. Different depths produce symmetrical, rotating lobes, with the rotation angle exhibiting an almost linear relationship with depth. This indicates that the angle between the lobes can be used to accurately estimate depth. The detection range is determined by the depth range over which the rotation angle changes by 180$^\circ$, and it can be extended by adjusting the target depth range during optimization. However, this comes at the cost of reduced spatial resolution.

\subsection{Experimental Characterization of DOF Performance} 

\begin{figure}
    \centering
    \includegraphics[width=\linewidth]{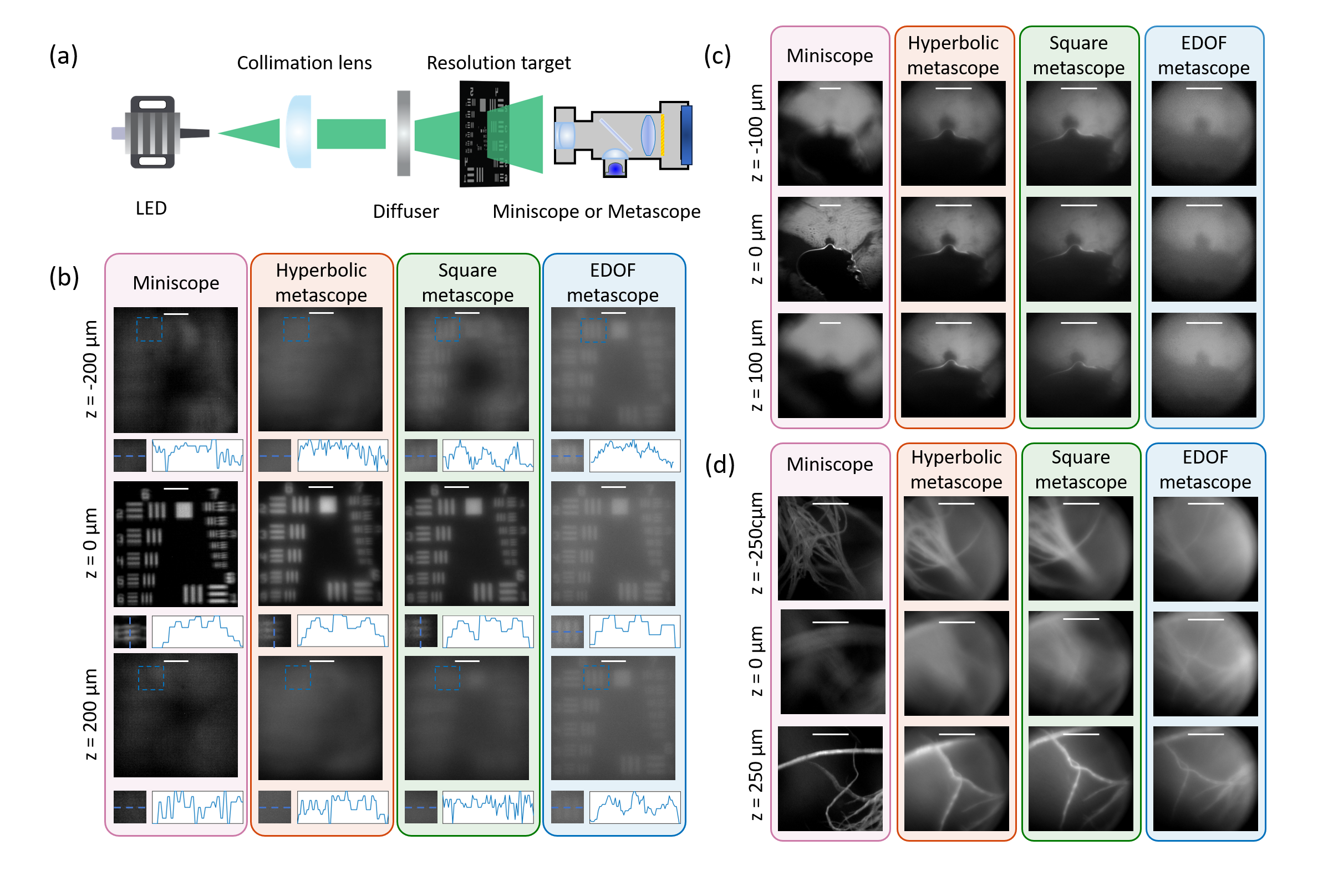}
    \caption{Performance Characterization of the Metascope (a) Schematic of the resolution imaging setup, consisting of an LED, collimation lens, diffuser, resolution target, and either a miniscope or metascope. For subsequent biological-sample imaging, the internal LED of the miniscope was used instead. (b) Resolution target images acquired with the miniscope and the hyperbolic, square, and EDOF metascopes (left to right) at axial positions of –200 \si{\micro\metre}, 0 \si{\micro\metre}, and 200 \si{\micro\metre}. Insets show intensity profiles across the smallest resolvable features. Blue dashed boxes mark the selected regions when visible. (c) Mouse kidney images acquired at axial positions of -100 \si{\micro\metre}, 0 \si{\micro\metre}, and 100 \si{\micro\metre}. (d) Multilayer fibrous sample images acquired at axial positions of –250 \si{\micro\metre}, 0 \si{\micro\metre}, and 250 \si{\micro\metre}. Scale bars: (b) 50 \si{\micro\metre}; (c, d) 200 \si{\micro\metre}.}
    \label{fig3}
\end{figure}

We then integrated the metalens with the miniscope, by placing the meta-optic at the miniscope entrance. We secured the backside of the metalens substrate to the objective module body using double-sided tape, instead of assembling the objective module body with achromatic lenses and a spacer. The assembly procedure for the remaining components followed the steps outlined in \cite{daniel_aharoni_2023_7844004}. Double-sided tape was chosen to facilitate easy swapping between different metalenses during characterization. However, the metalens could also be placed inside the module using 3D-printed mounts. Importantly, the TTL of the objective module can be reduced from 6.7 mm to 2.5 mm, and the WD can be extended from 0.7 mm to 2 mm by using the metalens, as shown in the inset of Figure~\ref{fig1}b. This would result in a significantly lower torque on the animal’s head. We first characterized the performance of different systems using a USAF 1951 resolution test target illuminated by a green LED (Thorlabs M530F1), whose center wavelength closely matches that of typical fluorophores. To simulate realistic imaging conditions, a diffuser was placed directly behind the resolution target. A schematic of the experimental setup is shown in Figure~\ref{fig3}a. For the resolution-target measurements, an external LED illumination setup was used because the target does not fluoresce. For subsequent biological-sample imaging, the internal LED illumination was used to excite fluorescence instead. The resolution target was mounted on an XYZ translation stage (Thorlabs PT3/M) to precisely adjust its axial position relative to the metascope, enabling image acquisition at both the focal plane and defocused planes. Negative defocus values are defined as object positions closer to the metascope, whereas positive values correspond to displacements farther away from it.

We captured images of the resolution target using different systems at positions ranging from $-200$~\si{\micro\metre} to 200~\si{\micro\metre}, shown in Figure~\ref{fig3}b. The columns from left to right show images acquired using the miniscope (refractive lens), followed by the metascopes equipped with the hyperbolic metalens, square metalens, and EDOF metalens, respectively (hereafter referred to as the “hyperbolic metascope”, “square metascope”, and “EDOF metascope”, respectively). All images presented here and in subsequent figures were processed as described in the Materials and Methods section. For each image, the smallest resolvable feature was identified, and a one-dimensional intensity profile was extracted across it to assess contrast, as shown in the insets in Figure~\ref{fig3}b. As designed, the EDOF metascope exhibits the largest DOF, maintaining consistent resolving capability across an axial range of ±200 \si{\micro\metre}. At axial positions of –200 \si{\micro\metre}, 0 \si{\micro\metre}, and 200 \si{\micro\metre}, it resolves features with line widths of approximately 6.96 \si{\micro\metre}, 4.38 \si{\micro\metre}, and 6.96 \si{\micro\metre}, respectively, corresponding to spatial resolutions of ~13.92 \si{\micro\metre} (71.8 lp/mm), ~8.76 \si{\micro\metre} (114.0 lp/mm), and ~13.92 \si{\micro\metre} (71.8 lp/mm). The square metascope exhibits an intermediate DOF, resolving features with line widths of approximately 6.96 \si{\micro\metre} (~13.92 \si{\micro\metre} resolution, 71.8 lp/mm) at –200 \si{\micro\metre}, but failing to resolve comparable features at 200~\si{\micro\metre}. In contrast, both the miniscope and the hyperbolic metascope could resolve fine feature lines only at the focal plane. At focus, the miniscope, hyperbolic metascope, and square metascope resolved features with line widths of approximately 2.76 \si{\micro\metre}, 2.76 \si{\micro\metre}, and 2.46 \si{\micro\metre}, respectively, corresponding to spatial resolutions of ~5.52 \si{\micro\metre} (181.0 lp/mm), ~5.52 \si{\micro\metre} (181.0 lp/mm), and ~4.92 \si{\micro\metre} (203.2 lp/mm). Even though the miniscope exhibits the clearest on-axis imaging, its field-dependent aberrations make it more difficult to distinguish features at the edge of the FOV, which highlights one of the advantages of our designed systems. Additional resolution target images and a summary table of the spatial resolution achieved by different systems at various axial depths are provided in Figure S5. Even in the PSF characterization section, the hyperbolic metalens exhibits a DOF comparable to that of the refractive lens under monochromatic illumination at 530 nm. However, the effective bandwidth is influenced by the illumination conditions and the built-in filters. As a result, the metascope shows an increased DOF compared with the miniscope, which can be attributed to the strong chromatic aberrations introduced by the metalens, as discussed in the Supporting Information.

Next, we demonstrate the applicability of our approach to biologically relevant samples, specifically a mouse kidney and a multilayer fibrous sample, to illustrate scenarios in which different DOF are required. The details of the sample preparation process can be found in the Materials and Methods section. We first captured images of the mouse kidney at axial positions of -100~\si{\micro\metre}, 0~\si{\micro\metre}, and 100~\si{\micro\metre}. The results are shown in Figure~\ref{fig3}c. For the miniscope, fine structural details are clearly resolved at the focal plane ($z = 0$~\si{\micro\metre}). However, as the imaging depth deviates from the focal plane, the image quality deteriorates rapidly. In contrast, the hyperbolic and square metascopes exhibit reduced clarity at the focal plane but preserve recognizable structural features over a broader axial range, with visible details maintained even at $z = \pm 100$~\si{\micro\metre}. This demonstrates an larger DOF for the metascope designs, which is beneficial for volumetric imaging applications. The EDOF metascope, although also exhibiting an improved DOF compared with the miniscope, suffers from low image contrast, making it less suitable for applications that require mid-range or high-resolution imaging across depth. Overall, these results highlight the substantially improved DOF of the metascope designs relative to the miniscope.

While the hyperbolic and square metascopes outperform the miniscope in terms of depth performance, their DOF remains limited, and in many cases, single-shot capture of information across a large volume rather than a single plane is essential, necessitating optics with extended DOF. This modality is highly valuable for 3D reconstruction of biological samples\cite{cao2023optical,barulin2025axially}, analysis of layered structures\cite{zhao2022flexible}, and generates more data than single plane captures. The EDOF metalens can capture information information across a larger DOF by optimizing the intensity distribution to accommodate specific application requirements. To evaluate the multi-plane imaging capability of the EDOF metascope, we prepared a multilayer synthetic fibrous sample by affixing commercial lens tissue to both sides of a 500~\si{\micro\metre} thick glass slide. The intertwined cellulose fibers, whose diameters range from a few micrometers to several hundred micrometers, form two spatially separated micro-textured layers. This configuration serves as a representative case for imaging applications where resolution demands are modest but a large DOF is essential. We captured images of the fibrous tissue sample at depths ranging from -250~\si{\micro\metre} to 250~\si{\micro\metre}, shown in Figure~\ref{fig3}d. While the structure remains clearly visible at the depths of ±250 \si{\micro\metre} for the miniscope, hyperbolic metascope, and square metascope, distinguishing the two layers becomes challenging at the middle position (depth 0). In comparison, although the EDOF metascope exhibits lower raw imaging quality due to pronounced aberrations, it still allows the overlapping structures from both layers to be observed at the middle position. For example, the tapered fiber segment oriented toward the lower right originates from the -250~\si{\micro\metre} layer, whereas the branching, tree-like structure visible in the lower-right region arises from the +250~\si{\micro\metre} layer. This superposition illustrates the multi-plane imaging capability enabled by the extended axial response of the EDOF design.

Our results demonstrate that some configurations of the metascope designs provide an increased DOF, in particular as demonstrated for the fibrous sample and the beads sample. Some other experiments are limited to effectively two-dimensional samples with only gradual axial variations. We note that for volumetric imaging, one must consider the inherent trade-off between DOF and spatial resolution, since signals from multiple axial planes can overlap. In principle, the DOF of our metalens systems can be adjusted by modifying the design parameters, as discussed in the Supporting Information. In addition to this dimensionality limitation, the biological samples in this study primarily serve to illustrate the comparative imaging behavior of the different lens systems rather than to resolve detailed morphology. Achieving high-contrast volumetric imaging or in-vivo demonstrations will require improvements in sample contrast, FOV, and imaging area, which we plan to explore in future work.

\subsection{Experimental Characterization of FOV Performance}

\begin{figure}
    \centering
    \includegraphics[width=\linewidth]{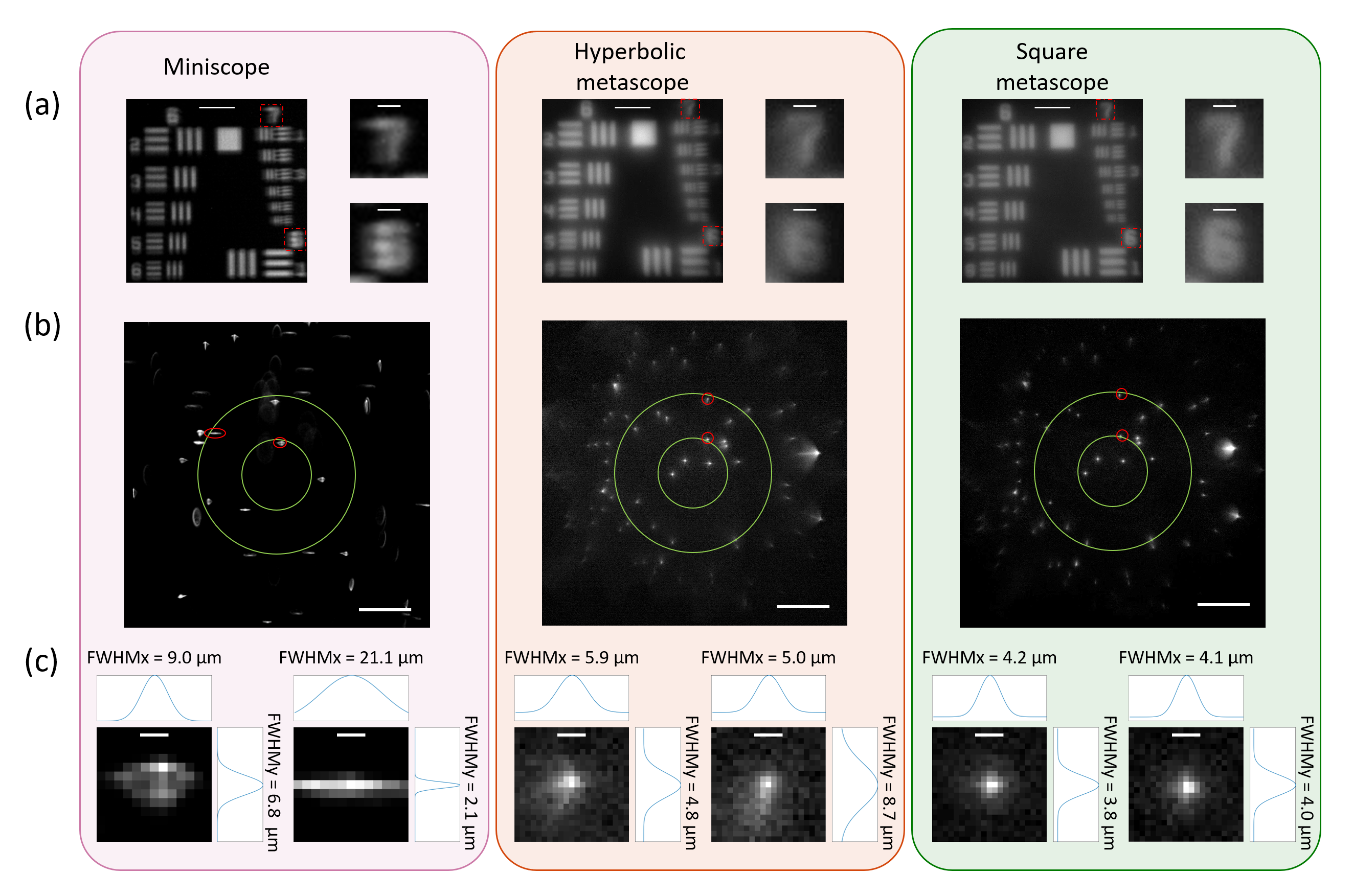}
    \caption{Comparison of the imaging performance of different systems for resolving resolution targets and micron-scale fluorescent beads. (a) Measured images of resolution targets acquired using three imaging systems: the miniscope (left), the hyperbolic metascope (center), and the square metascope (right). Representative regions near the periphery are highlighted by red rectangles to compare the aberrations. Scale bars: 50~\si{\micro\metre} (full-field images); 10~\si{\micro\metre} (insets). (b) Measured images of 1.9 \si{\micro\metre} fluorescent beads acquired using the same imaging systems. The overlaid green circles indicate identical FOV across all images for direct comparison. Scale bars, 100 \si{\micro\metre}. (c) PSF cross-sections corresponding to the circled structures in (b). Each subpanel shows the raw PSF image and its fitted Gaussian intensity profiles along the horizontal and vertical directions. FWHM values are annotated in each subpanel. Scale bars, 5 \si{\micro\metre}.}
    \label{fig4}
\end{figure}

In Figure~\ref{fig3}b, we show that the square metascope can resolve smaller features than the miniscope, demonstrating the superior performance of the square metascope when a large FOV is required. We additionally highlight the numbers at the periphery, as shown in Figure~\ref{fig4}a. For the miniscope, strong astigmatism is observed, and the number 6 is significantly stretched in the horizontal direction. In contrast, for the square metascope, the imaging appears blurry but the number remains legible. We further acquired images of the 1.9~\si{\micro\metre} fluorescent beads, which are distributed approximately within $\pm$50~\si{\micro\metre} along the detection plane, using the miniscope as well as the hyperbolic and square metascopes. The EDOF metascope exhibited substantial field aberrations, as shown in the earlier PSF characterization and resolution target measurements, which made quantitative analysis of the bead imaging unreliable; therefore, it was excluded from this section. Because the miniscope has a smaller magnification, we cropped the images acquired by the miniscope so that the same region in the sample corresponds to an equivalent FOV in the object plane as in the metascope images and the results are shown in Figure~\ref{fig4}b. Several beads appear elongated along either the horizontal or vertical direction, suggesting the presence of astigmatism in the imaging system. In contrast, for the hyperbolic and square metascopes, the bead shapes remain approximately circular, although slight coma aberration can be observed at large field positions. Additionally, due to their larger DOF, more beads can be discerned using the hyperbolic and square metascopes. We also capture the image ranging from -50~\si{\micro\metre} to 50~\si{\micro\metre}, as shown in the Figure S7. For the hyperbolic and square metalenses, the images remain relatively consistent, and the beads remain clearly visible throughout the axial range. In contrast, for the miniscope, due to the limited DOF, the beads appear and disappear at different depths, as only those whose focal positions are close to the detection plane can be clearly visualized. The large DOF and reduced aberrations of our system facilitate the identification of fine features, highlighting the advantages of our systems demonstrated in the previous section.

To better compare the resolution, we encircled two isolated spots that correspond to the same FOV across the different systems, as outlined in Figure~\ref{fig4}b. The FWHM was determined by fitting an elliptical Gaussian function to the circled PSFs. The PSF cross-sections and the corresponding fitted elliptical Gaussian intensity profiles along the horizontal and vertical directions are shown in Figure~\ref{fig4}c. For the miniscope, we selected two representative cases: one with the bead located at the tangential focal plane, exhibiting elongation along the horizontal direction while remaining sharply focused along the vertical axis; and another at an intermediate axial plane, where the elongation appears approximately symmetric in both directions. These represent two scenarios where the resolution along one or both dimensions plays a critical role. The calculated FWHMs in the horizontal and vertical directions are annotated in Figure~\ref{fig4}c. Compared to the miniscope, the hyperbolic and square metascopes exhibit smaller and more symmetric PSFs across the FOV, indicating improved aberration correction and resolution uniformity. For the miniscope, although it can achieve very high resolution in one direction, its strong astigmatism significantly degrades performance when high resolution in all directions is required. In addition, strong field curvature can degrade the resolution in both directions, as evidenced by the second set of beads exhibiting reduced resolution along both axes. In contrast, the square metascope provides better overall resolution across both directions. The hyperbolic metascope also exhibits good resolution in both directions; however, the strong noise present in its images reduces the contrast. 

The pronounced astigmatism may result from a misalignment in the refractive objective module, given that no similar aberration is observed in the metascope configuration. However, a more quantitative analysis of this effect may be necessary to evaluate the imaging performance of the entire system, rather than attributing it solely to the objective module, which is beyond the scope of this work. We also note that this represents another advantage of our system: it reduces reliance on a cascaded series of refractive components, thereby improving robustness against misalignment—particularly in scenarios where additional refractive elements are required to further enhance imaging quality \cite{guo2023miniscope}. It is also important to emphasize that the achievable FOV in the miniscope system is primarily determined by the system magnification, given a fixed sensor size. Since our metascope design employs a higher magnification (5×) than the commercial miniscope (2.9×), the resulting FOV is reduced to approximately 0.6 mm × 0.6 mm, compared with the typical~1 mm × 1 mm FOV of conventional miniscope systems. However, modifying the focal length of a metalens is relatively straightforward, as it only requires updating the phase profile or adjusting the optimization parameters. Further improvement of the resolution and reduction of aberrations can be achieved by optimizing the phase profile to incorporate the aberrations of the entire imaging system, rather than considering only the objective module as was done in this work. A more detailed system-level optimization is provided in the Supporting Information.

\subsection{Experimental Characterization of Depth Sensing Performance}

\begin{figure}
    \centering
     \includegraphics[width=\linewidth]{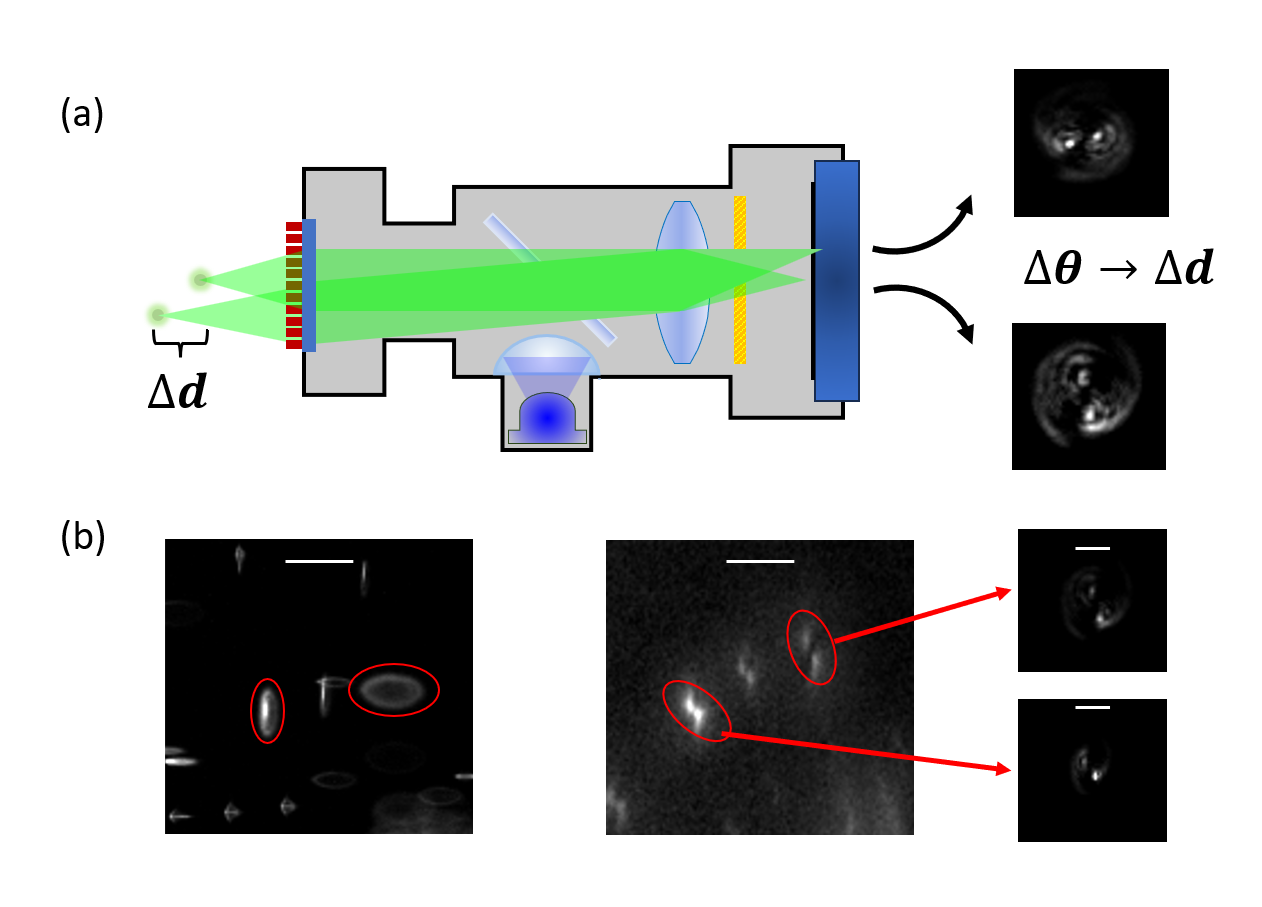}
    \caption{Depth sensing using the DH metascope (a) Schematic diagram illustrating the working principle of depth sensing. Depth information is extracted from the angular differences observed in the measured PSFs. For clarity, the light emitted by the LED is not depicted. (b) Captured images of 1.9~\si{\micro\metre} fluorescent beads acquired using the miniscope and the DH metascope. The left panel shows the image captured by the miniscope, while the right panels present images acquired by the DH metascope and the corresponding calibration PSFs with similar rotation angles. The red circles highlight representative PSFs used for depth estimation. Scale bars: 50~\si{\micro\metre} for wide-field images; 20~\si{\micro\metre} for PSF images.}
    \label{fig5}
\end{figure}

The DH metalens is characterized by two lobes that continuously rotate with change in focal plane, thus depth information can be encoded through the rotation angle, facilitating precise depth sensing. A schematic diagram of the working principle is shown in Figure~\ref{fig5}a. As shown in the optical performance characterization section, the relationship between depth and rotation angle is approximately linear. However, for practical applications in which the DH metalens is integrated with the miniscope, the system aberrations must be taken into account to enable more precise measurements. To address this issue, we calibrated the relationship between the rotation angle and axial depth. The experimental procedure is described in detail in the Supporting Information, and the corresponding calibration results, illustrating the depth as a function of rotation angle, are provided in the accompanying Supporting video.

During the experiment, we found that only beads near the central field exhibited clearly observable two rotating lobes at different depths, whereas for the other beads, either the intensity was too weak or the two lobes were difficult to distinguish. This arises from the limited FOV of the designed DH metalens, as the design was optimized exclusively for normal incidence. An analysis of the impact of the AOI on the performance is provided in Supporting Information. We observe that the PSF shape of the DH metalens is strongly influenced by off-axis aberrations. For more accurate results, calibration accounting for the AOI may also be necessary, as discussed in \cite{colburn2020metasurface}. As demonstrated in several prior works\cite{hao2024single,kim2025rapid,colburn2020metasurface}, DH metalens can enable depth-encoded 3D imaging. For miniscope applications, undistorted images can be recovered through deconvolution, or alternatively by incorporating an additional imaging path. A lens-array architecture that integrates both DH and hyperbolic metalenses is also feasible. As discussed in the Supporting Information, further system-level optimization could enhance the DH metascope and unlock its full 3D imaging capability. Here, we only present the prototype, and further improvements such as increasing the FOV, detection range, or resolution can be pursued in future work. 

For the ground truth, we used the miniscope to measure their relative axial positions by taking advantage of its limited DOF, which facilitated precise focusing. We selected the position where both beads are clearly visible, as shown in Figure~\ref{fig5}b. We then moved the two beads to the center and determined their axial positions, respectively. The left bead was located 50 \si{\micro\metre} farther away, which serves as the ground truth in the subsequent discussion. We used the procedure described in the Supporting Information to calculate the depth difference between the two beads. First, we adjusted the focal length of the electrowetting lens to position both beads within the negative defocus region, as shown in Figure~\ref{fig5}b. We observed that the distance between the two beads is below the detection range of each defocus section, allowing them to appear simultaneously in either the positive or negative defocus positions. However, if their depth difference exceeds the detection range of a single section, each bead can still be detected separately. The angles of two lobes for two beads are 50.33$^\circ$ and 70.44$^\circ$ with respect to the negative $x$ axis, respectively. We then identified the calibration PSFs in the negative defocus region with similar rotation angles. For depths of –55~\si{\micro\metre} and –100~\si{\micro\metre}, the corresponding angles are 50.46$^\circ$ and 69.90$^\circ$, respectively, as shown in Figure~\ref{fig5}(b). Thus, we estimate the depth difference to be 45~\si{\micro\metre}, which is close to the ground truth. The detection resolution can be increased by reducing the increment in the calibration steps. Additionally, the calculation process could be accelerated by automatically detecting isolated beads and deconvolving with the calibration PSF.

\section{CONCLUSIONS}

We demonstrate the integration of an extensively studied single-layer metalens with an open-source miniscope. This approach combines the advantages of the metalens, such as its compact footprint and multi-functionality, with the strengths of the miniscope, including its capability for in vivo imaging and high frame rate performance. We demonstrate the integration of various metalenses, including hyperbolic, square, EDOF, and DH metalenses. Our results highlight imaging capabilities with an extended DOF, a wide FOV, and precise depth sensing. This integration paves the way for compact and versatile imaging systems suitable for a wide range of applications. Furthermore, some other functionalities that have been explored in the metasurface community could also be integrated through careful design. For instance, specially designed meta-optics can be employed for edge detection under specific illumination conditions and contour extraction in biological imaging. Polarization sensitive meta-optics could allow for polarimetry, or stereo imaging. In addition, hyperspectral imaging could be realized by exploiting the strong chromatic dispersion of metasurfaces or by adopting lens array configurations \cite{audhkhasi2025single}. We emphasis that the integration of system is easy as only the replacement of the objective lens is required. Also, we envision the replacement of the tube lens \cite{yanny2020miniscope3d} to provide more design freedom and further reduce the system's complexity.

Another important advantage of the metalens is its ability to provide a large working distance in the system. For the metalens, the back focal length (BFL)—defined as the distance from the image plane to the last surface of the lens system—is approximately equal to the effective focal length (EFL), since the metalens has negligible physical thickness. However, in the case of objective lenses used in miniscopes, the BFL is typically much smaller than the EFL. This limits their applicability in scenarios that require a large working distance, particularly in in vivo imaging.

While we observe that the imaging quality of our system remains significantly below that of the original miniscope—an expected outcome, as similar phenomena have been noted in free-space setups—there is substantial room for improvement. For example, our current design optimizes only the objective lens itself, while neglecting aberrations introduced by other components in the overall imaging system. Optimizing the phase mask of the meta-optics within the context of the entire imaging system could lead to improved overall performance. In addition, computational techniques such as deconvolution or neural networks may be employed to further enhance image quality. Moreover, adopting an end-to-end design strategy presents a promising direction for future work, with the potential to significantly advance imaging performance.  Although meta-optics currently require specialized expertise and fabrication processes, recent progress in scalable manufacturing and automated design frameworks is rapidly reducing this barrier (see Supporting Information).

\section{MATERIALS AND METHODS}

\subsection{Fabrication}

For fabrication, a 800 nm silicon nitride film was first deposited onto a 500~\si{\micro\metre} double-side-polished fused silica wafer using plasma-enhanced chemical vapor deposition (PECVD). The wafer was then diced into 1.5 cm pieces using a Disco Wafer Dicer. For spin coating, a positive resist (ZEP 520A) was applied to the chip and baked at 180°C for 3 minutes. Subsequently, a discharge chemical (DisCharge H2O) was spun onto the chip to mitigate charging effects. The resist layer was then patterned using a 100 kV electron beam (JEOL JBX6300FS) at a dose of $\sim$~260~\si{\micro\C}$ \:\mathrm{cm}^{-2}$ and developed in Amyl Acetate for 2 minutes. A $\sim$~60 nm aluminum oxide thin film was deposited as a hard mask using electron beam evaporation. After an overnight lift-off process in NMP, the chip was etched using an inductively coupled reactive ion etcher (Oxford PlasmaLab System 100).

\subsection{Image Processing and Background Correction}

All biological imaging captured using miniscope or metascopes underwent only minimal and standardized pre-processing steps to ensure a fair comparison across different optical systems. During fluorescence imaging, the exposure settings (frame rate, gain, and illumination intensity) were adjusted in each system to ensure clear, non-saturated images. The stray light originating from internal reflections in the internal systems introduces a spatially varying background. To correct for this effect, a background frame was recorded under identical exposure conditions by translating the sample to an empty region. This background image captures only the system-induced stray light and sensor offset. The final image was obtained using:

\begin{equation}
    \mathrm{I_{corrected} = I_{raw}-I_{background}}
\end{equation}

After background subtraction, each image was linearly normalized to its full dynamic range.

\subsection{Biological Sample Preparation}

All protocols and methods involving animals in this work were approved by the Institutional Animal Care and Use Committee at University of Washington. Four-month old C57BL/6 male mice were anesthetized using an isoflurane/oxygen mixture followed by cardiac perfusion with 1× PBS and 4$\%$ paraformaldehyde (PFA) for 5 min each. Kidneys were collected and fixed in 4$\%$ PFA in 1× PBS for 1 to 6 hours. The kidneys were then washed with 1× PBS and sliced using a vibratome to 100-\si{\micro\metre} thick sections. Kidney sections were stained with Sybr Green I dye (S7563, Invitrogen) to label the nuclei and mounted on a glass slide with Prolong Diamond Antifade (P36965, Thermo Fisher Scientific).  

Fluorescent polystyrene beads of 1.9~\si{\micro\metre} diameter (Fluospheres, Thermo Fisher), excitable with a 488 nm laser, were mounted with ProLong Diamond Antifade in a channel of fixed height created by sandwiching a parafilm spacer between a coverslip and a glass slide.

A multilayer fibrous sample was prepared by attaching lens tissue (MC-5, Thorlabs) to both sides of a 500~\si{\micro\metre}-thick fused silica chip (University Wafer). The lens tissue sheets were carefully trimmed to match the dimensions of the chip, and a yellow highlighter was gently applied to the surface to introduce fluorescent labeling. The resulting sample was then mounted onto a standard glass microscope slide for subsequent imaging.

\subsection{Author Contributions}

A.M. and J.F. conceived the idea. C.P. and N.Z.X. prepared the biological sample. Z.H.Z. and J.F. performed the simulations, design, and fabrication of the metalenses. J.F., P.C., S.C., and N.Z.X. provided suggestions for the experiments. Z.H.Z. and J.F. characterized the optical performance of the metalenses. Z.H.Z. and K.K. performed the characterization of the miniscope and metascope. Z.H.Z. analysed the data with input from S.C., P.C., and J.F.. K.B., A.M. and J.F supervised the project. Z.H.Z. and J.F. wrote the main parts of the manuscript. All authors contributed to writing the manuscript, discussing the results, and providing feedback on the manuscript.

\subsection{Notes}

A.M. and K.B. are confounders of Tunoptix, which aims to commercialize meta-optics technology. The remaining authors declare no competing interests.

\begin{acknowledgement}

The work is supported by Washington Research Foundation. Part of this work was conducted at the Washington Nanofabrication Facility / Molecular Analysis Facility, a National Nanotechnology Coordinated Infrastructure (NNCI) site at the University of Washington with partial support from the National Science Foundation via awards NNCI-1542101 and NNCI-2025489.
This research was supported by an appointment to the Intelligence Community Postdoctoral Research Fellowship Program at the University of Washington administered by Oak Ridge Institute for Science and Education (ORISE) through an interagency agreement between the U.S. Department of Energy and the Office of the Director of National Intelligence (ODNI).
We gratefully acknowledge professor Joshua Vaughan, Department of Chemistry, University of Washington, for providing expert assistance in the preparation and handling of biological specimens used in this study.

\end{acknowledgement}

\bibliography{main_text}

\providecommand{\latin}[1]{#1}
\makeatletter
\providecommand{\doi}
  {\begingroup\let\do\@makeother\dospecials
  \catcode`\{=1 \catcode`\}=2 \doi@aux}
\providecommand{\doi@aux}[1]{\endgroup\texttt{#1}}
\makeatother
\providecommand*\mcitethebibliography{\thebibliography}
\csname @ifundefined\endcsname{endmcitethebibliography}  {\let\endmcitethebibliography\endthebibliography}{}
\begin{mcitethebibliography}{56}
\providecommand*\natexlab[1]{#1}
\providecommand*\mciteSetBstSublistMode[1]{}
\providecommand*\mciteSetBstMaxWidthForm[2]{}
\providecommand*\mciteBstWouldAddEndPuncttrue
  {\def\EndOfBibitem{\unskip.}}
\providecommand*\mciteBstWouldAddEndPunctfalse
  {\let\EndOfBibitem\relax}
\providecommand*\mciteSetBstMidEndSepPunct[3]{}
\providecommand*\mciteSetBstSublistLabelBeginEnd[3]{}
\providecommand*\EndOfBibitem{}
\mciteSetBstSublistMode{f}
\mciteSetBstMaxWidthForm{subitem}{(\alph{mcitesubitemcount})}
\mciteSetBstSublistLabelBeginEnd
  {\mcitemaxwidthsubitemform\space}
  {\relax}
  {\relax}

\bibitem[Lichtman and Conchello(2005)Lichtman, and Conchello]{lichtman2005fluorescence}
Lichtman,~J.~W.; Conchello,~J.-A. Fluorescence microscopy. \emph{Nature methods} \textbf{2005}, \emph{2}, 910--919\relax
\mciteBstWouldAddEndPuncttrue
\mciteSetBstMidEndSepPunct{\mcitedefaultmidpunct}
{\mcitedefaultendpunct}{\mcitedefaultseppunct}\relax
\EndOfBibitem
\bibitem[Rao \latin{et~al.}(2007)Rao, Dragulescu-Andrasi, and Yao]{rao2007fluorescence}
Rao,~J.; Dragulescu-Andrasi,~A.; Yao,~H. Fluorescence imaging in vivo: recent advances. \emph{Current opinion in biotechnology} \textbf{2007}, \emph{18}, 17--25\relax
\mciteBstWouldAddEndPuncttrue
\mciteSetBstMidEndSepPunct{\mcitedefaultmidpunct}
{\mcitedefaultendpunct}{\mcitedefaultseppunct}\relax
\EndOfBibitem
\bibitem[Barretto \latin{et~al.}(2009)Barretto, Messerschmidt, and Schnitzer]{barretto2009vivo}
Barretto,~R.~P.; Messerschmidt,~B.; Schnitzer,~M.~J. In vivo fluorescence imaging with high-resolution microlenses. \emph{Nature methods} \textbf{2009}, \emph{6}, 511--512\relax
\mciteBstWouldAddEndPuncttrue
\mciteSetBstMidEndSepPunct{\mcitedefaultmidpunct}
{\mcitedefaultendpunct}{\mcitedefaultseppunct}\relax
\EndOfBibitem
\bibitem[Ghosh \latin{et~al.}(2011)Ghosh, Burns, Cocker, Nimmerjahn, Ziv, Gamal, and Schnitzer]{ghosh2011miniaturized}
Ghosh,~K.~K.; Burns,~L.~D.; Cocker,~E.~D.; Nimmerjahn,~A.; Ziv,~Y.; Gamal,~A.~E.; Schnitzer,~M.~J. Miniaturized integration of a fluorescence microscope. \emph{Nature methods} \textbf{2011}, \emph{8}, 871--878\relax
\mciteBstWouldAddEndPuncttrue
\mciteSetBstMidEndSepPunct{\mcitedefaultmidpunct}
{\mcitedefaultendpunct}{\mcitedefaultseppunct}\relax
\EndOfBibitem
\bibitem[Stamatakis \latin{et~al.}(2021)Stamatakis, Resendez, Chen, Favero, Liang-Guallpa, Nassi, Neufeld, Visscher, and Ghosh]{stamatakis2021miniature}
Stamatakis,~A.~M.; Resendez,~S.~L.; Chen,~K.-S.; Favero,~M.; Liang-Guallpa,~J.; Nassi,~J.~J.; Neufeld,~S.~Q.; Visscher,~K.; Ghosh,~K.~K. Miniature microscopes for manipulating and recording in vivo brain activity. \emph{Microscopy} \textbf{2021}, \emph{70}, 399--414\relax
\mciteBstWouldAddEndPuncttrue
\mciteSetBstMidEndSepPunct{\mcitedefaultmidpunct}
{\mcitedefaultendpunct}{\mcitedefaultseppunct}\relax
\EndOfBibitem
\bibitem[Chen \latin{et~al.}(2022)Chen, Tian, and Kong]{chen2022advances}
Chen,~K.; Tian,~Z.; Kong,~L. Advances of optical miniscopes for in vivo imaging of neural activity in freely moving animals. \emph{Frontiers in Neuroscience} \textbf{2022}, \emph{16}, 994079\relax
\mciteBstWouldAddEndPuncttrue
\mciteSetBstMidEndSepPunct{\mcitedefaultmidpunct}
{\mcitedefaultendpunct}{\mcitedefaultseppunct}\relax
\EndOfBibitem
\bibitem[Aharoni and Hoogland(2019)Aharoni, and Hoogland]{aharoni2019circuit}
Aharoni,~D.; Hoogland,~T.~M. Circuit investigations with open-source miniaturized microscopes: past, present and future. \emph{Frontiers in cellular neuroscience} \textbf{2019}, \emph{13}, 141\relax
\mciteBstWouldAddEndPuncttrue
\mciteSetBstMidEndSepPunct{\mcitedefaultmidpunct}
{\mcitedefaultendpunct}{\mcitedefaultseppunct}\relax
\EndOfBibitem
\bibitem[Guo \latin{et~al.}(2023)Guo, Blair, Sehgal, Sangiuliano~Jimka, Bellafard, Silva, Golshani, Basso, Blair, and Aharoni]{guo2023miniscope}
Guo,~C.; Blair,~G.~J.; Sehgal,~M.; Sangiuliano~Jimka,~F.~N.; Bellafard,~A.; Silva,~A.~J.; Golshani,~P.; Basso,~M.~A.; Blair,~H.~T.; Aharoni,~D. Miniscope-LFOV: a large-field-of-view, single-cell-resolution, miniature microscope for wired and wire-free imaging of neural dynamics in freely behaving animals. \emph{Science advances} \textbf{2023}, \emph{9}, eadg3918\relax
\mciteBstWouldAddEndPuncttrue
\mciteSetBstMidEndSepPunct{\mcitedefaultmidpunct}
{\mcitedefaultendpunct}{\mcitedefaultseppunct}\relax
\EndOfBibitem
\bibitem[Liberti \latin{et~al.}(2017)Liberti, Perkins, Leman, and Gardner]{liberti2017open}
Liberti,~W.~A.; Perkins,~L.~N.; Leman,~D.~P.; Gardner,~T.~J. An open source, wireless capable miniature microscope system. \emph{Journal of neural engineering} \textbf{2017}, \emph{14}, 045001\relax
\mciteBstWouldAddEndPuncttrue
\mciteSetBstMidEndSepPunct{\mcitedefaultmidpunct}
{\mcitedefaultendpunct}{\mcitedefaultseppunct}\relax
\EndOfBibitem
\bibitem[Prevedel \latin{et~al.}(2014)Prevedel, Yoon, Hoffmann, Pak, Wetzstein, Kato, Schr{\"o}del, Raskar, Zimmer, Boyden, \latin{et~al.} others]{prevedel2014simultaneous}
Prevedel,~R.; Yoon,~Y.-G.; Hoffmann,~M.; Pak,~N.; Wetzstein,~G.; Kato,~S.; Schr{\"o}del,~T.; Raskar,~R.; Zimmer,~M.; Boyden,~E.~S.; others Simultaneous whole-animal 3D imaging of neuronal activity using light-field microscopy. \emph{Nature methods} \textbf{2014}, \emph{11}, 727--730\relax
\mciteBstWouldAddEndPuncttrue
\mciteSetBstMidEndSepPunct{\mcitedefaultmidpunct}
{\mcitedefaultendpunct}{\mcitedefaultseppunct}\relax
\EndOfBibitem
\bibitem[Skocek \latin{et~al.}(2018)Skocek, N{\"o}bauer, Weilguny, Mart{\'\i}nez~Traub, Xia, Molodtsov, Grama, Yamagata, Aharoni, Cox, \latin{et~al.} others]{skocek2018high}
Skocek,~O.; N{\"o}bauer,~T.; Weilguny,~L.; Mart{\'\i}nez~Traub,~F.; Xia,~C.~N.; Molodtsov,~M.~I.; Grama,~A.; Yamagata,~M.; Aharoni,~D.; Cox,~D.~D.; others High-speed volumetric imaging of neuronal activity in freely moving rodents. \emph{Nature methods} \textbf{2018}, \emph{15}, 429--432\relax
\mciteBstWouldAddEndPuncttrue
\mciteSetBstMidEndSepPunct{\mcitedefaultmidpunct}
{\mcitedefaultendpunct}{\mcitedefaultseppunct}\relax
\EndOfBibitem
\bibitem[Yanny \latin{et~al.}(2020)Yanny, Antipa, Liberti, Dehaeck, Monakhova, Liu, Shen, Ng, and Waller]{yanny2020miniscope3d}
Yanny,~K.; Antipa,~N.; Liberti,~W.; Dehaeck,~S.; Monakhova,~K.; Liu,~F.~L.; Shen,~K.; Ng,~R.; Waller,~L. Miniscope3D: optimized single-shot miniature 3D fluorescence microscopy. \emph{Light: Science \& Applications} \textbf{2020}, \emph{9}, 171\relax
\mciteBstWouldAddEndPuncttrue
\mciteSetBstMidEndSepPunct{\mcitedefaultmidpunct}
{\mcitedefaultendpunct}{\mcitedefaultseppunct}\relax
\EndOfBibitem
\bibitem[Xue \latin{et~al.}(2020)Xue, Davison, Boas, and Tian]{xue2020single}
Xue,~Y.; Davison,~I.~G.; Boas,~D.~A.; Tian,~L. Single-shot 3D wide-field fluorescence imaging with a Computational Miniature Mesoscope. \emph{Science advances} \textbf{2020}, \emph{6}, eabb7508\relax
\mciteBstWouldAddEndPuncttrue
\mciteSetBstMidEndSepPunct{\mcitedefaultmidpunct}
{\mcitedefaultendpunct}{\mcitedefaultseppunct}\relax
\EndOfBibitem
\bibitem[Greene \latin{et~al.}(2023)Greene, Xue, Alido, Matlock, Hu, Kili{\c{c}}, Davison, and Tian]{greene2023pupil}
Greene,~J.; Xue,~Y.; Alido,~J.; Matlock,~A.; Hu,~G.; Kili{\c{c}},~K.; Davison,~I.; Tian,~L. Pupil engineering for extended depth-of-field imaging in a fluorescence miniscope. \emph{Neurophotonics} \textbf{2023}, \emph{10}, 044302--044302\relax
\mciteBstWouldAddEndPuncttrue
\mciteSetBstMidEndSepPunct{\mcitedefaultmidpunct}
{\mcitedefaultendpunct}{\mcitedefaultseppunct}\relax
\EndOfBibitem
\bibitem[Scherrer \latin{et~al.}(2023)Scherrer, Lynch, Zhang, and Fee]{scherrer2023optical}
Scherrer,~J.~R.; Lynch,~G.~F.; Zhang,~J.~J.; Fee,~M.~S. An optical design enabling lightweight and large field-of-view head-mounted microscopes. \emph{Nature Methods} \textbf{2023}, \emph{20}, 546--549\relax
\mciteBstWouldAddEndPuncttrue
\mciteSetBstMidEndSepPunct{\mcitedefaultmidpunct}
{\mcitedefaultendpunct}{\mcitedefaultseppunct}\relax
\EndOfBibitem
\bibitem[Zhao \latin{et~al.}(2025)Zhao, Guo, Xie, Chen, Golshani, and Aharoni]{zhao2025minixl}
Zhao,~P.; Guo,~C.; Xie,~M.; Chen,~L.; Golshani,~P.; Aharoni,~D. MiniXL: An open-source, large field-of-view epifluorescence miniscope enabling single-cell resolution and multi-region imaging in mice. \emph{Science Advances} \textbf{2025}, \emph{11}, eads4995\relax
\mciteBstWouldAddEndPuncttrue
\mciteSetBstMidEndSepPunct{\mcitedefaultmidpunct}
{\mcitedefaultendpunct}{\mcitedefaultseppunct}\relax
\EndOfBibitem
\bibitem[Yang \latin{et~al.}(2024)Yang, Guo, Hu, Xue, Li, and Tian]{yang2024wide}
Yang,~Q.; Guo,~R.; Hu,~G.; Xue,~Y.; Li,~Y.; Tian,~L. Wide-field, high-resolution reconstruction in computational multi-aperture miniscope using a Fourier neural network. \emph{Optica} \textbf{2024}, \emph{11}, 860--871\relax
\mciteBstWouldAddEndPuncttrue
\mciteSetBstMidEndSepPunct{\mcitedefaultmidpunct}
{\mcitedefaultendpunct}{\mcitedefaultseppunct}\relax
\EndOfBibitem
\bibitem[Adams \latin{et~al.}(2017)Adams, Boominathan, Avants, Vercosa, Ye, Baraniuk, Robinson, and Veeraraghavan]{adams2017single}
Adams,~J.~K.; Boominathan,~V.; Avants,~B.~W.; Vercosa,~D.~G.; Ye,~F.; Baraniuk,~R.~G.; Robinson,~J.~T.; Veeraraghavan,~A. Single-frame 3D fluorescence microscopy with ultraminiature lensless FlatScope. \emph{Science advances} \textbf{2017}, \emph{3}, e1701548\relax
\mciteBstWouldAddEndPuncttrue
\mciteSetBstMidEndSepPunct{\mcitedefaultmidpunct}
{\mcitedefaultendpunct}{\mcitedefaultseppunct}\relax
\EndOfBibitem
\bibitem[Adams \latin{et~al.}(2022)Adams, Yan, Wu, Boominathan, Gao, Rodriguez, Kim, Carns, Richards-Kortum, Kemere, \latin{et~al.} others]{adams2022vivo}
Adams,~J.~K.; Yan,~D.; Wu,~J.; Boominathan,~V.; Gao,~S.; Rodriguez,~A.~V.; Kim,~S.; Carns,~J.; Richards-Kortum,~R.; Kemere,~C.; others In vivo lensless microscopy via a phase mask generating diffraction patterns with high-contrast contours. \emph{Nature Biomedical Engineering} \textbf{2022}, \emph{6}, 617--628\relax
\mciteBstWouldAddEndPuncttrue
\mciteSetBstMidEndSepPunct{\mcitedefaultmidpunct}
{\mcitedefaultendpunct}{\mcitedefaultseppunct}\relax
\EndOfBibitem
\bibitem[Kuznetsov \latin{et~al.}(2024)Kuznetsov, Brongersma, Yao, Chen, Levy, Tsai, Zheludev, Faraon, Arbabi, Yu, \latin{et~al.} others]{kuznetsov2024roadmap}
Kuznetsov,~A.~I.; Brongersma,~M.~L.; Yao,~J.; Chen,~M.~K.; Levy,~U.; Tsai,~D.~P.; Zheludev,~N.~I.; Faraon,~A.; Arbabi,~A.; Yu,~N.; others Roadmap for optical metasurfaces. \emph{ACS photonics} \textbf{2024}, \emph{11}, 816--865\relax
\mciteBstWouldAddEndPuncttrue
\mciteSetBstMidEndSepPunct{\mcitedefaultmidpunct}
{\mcitedefaultendpunct}{\mcitedefaultseppunct}\relax
\EndOfBibitem
\bibitem[Hsiao \latin{et~al.}(2017)Hsiao, Chu, and Tsai]{hsiao2017fundamentals}
Hsiao,~H.-H.; Chu,~C.~H.; Tsai,~D.~P. Fundamentals and applications of metasurfaces. \emph{Small Methods} \textbf{2017}, \emph{1}, 1600064\relax
\mciteBstWouldAddEndPuncttrue
\mciteSetBstMidEndSepPunct{\mcitedefaultmidpunct}
{\mcitedefaultendpunct}{\mcitedefaultseppunct}\relax
\EndOfBibitem
\bibitem[Wei \latin{et~al.}(2021)Wei, Cao, Lin, Yuan, Somekh, and Jia]{wei2021varifocal}
Wei,~S.; Cao,~G.; Lin,~H.; Yuan,~X.; Somekh,~M.; Jia,~B. A varifocal graphene metalens for broadband zoom imaging covering the entire visible region. \emph{ACS nano} \textbf{2021}, \emph{15}, 4769--4776\relax
\mciteBstWouldAddEndPuncttrue
\mciteSetBstMidEndSepPunct{\mcitedefaultmidpunct}
{\mcitedefaultendpunct}{\mcitedefaultseppunct}\relax
\EndOfBibitem
\bibitem[Colburn and Majumdar(2019)Colburn, and Majumdar]{colburn2019simultaneous}
Colburn,~S.; Majumdar,~A. Simultaneous achromatic and varifocal imaging with quartic metasurfaces in the visible. \emph{ACS Photonics} \textbf{2019}, \emph{7}, 120--127\relax
\mciteBstWouldAddEndPuncttrue
\mciteSetBstMidEndSepPunct{\mcitedefaultmidpunct}
{\mcitedefaultendpunct}{\mcitedefaultseppunct}\relax
\EndOfBibitem
\bibitem[Wang \latin{et~al.}(2021)Wang, Intaravanne, Li, Han, Chen, Liu, Zhang, Li, and Chen]{wang2021metalens}
Wang,~R.; Intaravanne,~Y.; Li,~S.; Han,~J.; Chen,~S.; Liu,~J.; Zhang,~S.; Li,~L.; Chen,~X. Metalens for generating a customized vectorial focal curve. \emph{Nano Letters} \textbf{2021}, \emph{21}, 2081--2087\relax
\mciteBstWouldAddEndPuncttrue
\mciteSetBstMidEndSepPunct{\mcitedefaultmidpunct}
{\mcitedefaultendpunct}{\mcitedefaultseppunct}\relax
\EndOfBibitem
\bibitem[Lin \latin{et~al.}(2024)Lin, Geldmeier, Baleine, Yang, An, Pan, Rivero-Baleine, Gu, and Hu]{lin2024wide}
Lin,~H.-I.; Geldmeier,~J.; Baleine,~E.; Yang,~F.; An,~S.; Pan,~Y.; Rivero-Baleine,~C.; Gu,~T.; Hu,~J. Wide-Field-of-View, Large-Area Long-Wave Infrared Silicon Metalenses. \emph{ACS Photonics} \textbf{2024}, \emph{11}, 1943--1949\relax
\mciteBstWouldAddEndPuncttrue
\mciteSetBstMidEndSepPunct{\mcitedefaultmidpunct}
{\mcitedefaultendpunct}{\mcitedefaultseppunct}\relax
\EndOfBibitem
\bibitem[Wirth-Singh \latin{et~al.}(2025)Wirth-Singh, Fr{\"o}ch, Yang, Martin, Zheng, Zhang, Tanguy, Zhou, Huang, John, \latin{et~al.} others]{wirth2025wide}
Wirth-Singh,~A.; Fr{\"o}ch,~J.~E.; Yang,~F.; Martin,~L.; Zheng,~H.; Zhang,~H.; Tanguy,~Q.~T.; Zhou,~Z.; Huang,~L.; John,~D.~D.; others Wide field of view large aperture meta-doublet eyepiece. \emph{Light: Science \& Applications} \textbf{2025}, \emph{14}, 17\relax
\mciteBstWouldAddEndPuncttrue
\mciteSetBstMidEndSepPunct{\mcitedefaultmidpunct}
{\mcitedefaultendpunct}{\mcitedefaultseppunct}\relax
\EndOfBibitem
\bibitem[Lee \latin{et~al.}(2018)Lee, Hong, Hwang, Moon, Kang, Jeon, Kim, Jeong, and Lee]{lee2018metasurface}
Lee,~G.-Y.; Hong,~J.-Y.; Hwang,~S.; Moon,~S.; Kang,~H.; Jeon,~S.; Kim,~H.; Jeong,~J.-H.; Lee,~B. Metasurface eyepiece for augmented reality. \emph{Nature communications} \textbf{2018}, \emph{9}, 4562\relax
\mciteBstWouldAddEndPuncttrue
\mciteSetBstMidEndSepPunct{\mcitedefaultmidpunct}
{\mcitedefaultendpunct}{\mcitedefaultseppunct}\relax
\EndOfBibitem
\bibitem[Gopakumar \latin{et~al.}(2024)Gopakumar, Lee, Choi, Chao, Peng, Kim, and Wetzstein]{gopakumar2024full}
Gopakumar,~M.; Lee,~G.-Y.; Choi,~S.; Chao,~B.; Peng,~Y.; Kim,~J.; Wetzstein,~G. Full-colour 3D holographic augmented-reality displays with metasurface waveguides. \emph{Nature} \textbf{2024}, \emph{629}, 791--797\relax
\mciteBstWouldAddEndPuncttrue
\mciteSetBstMidEndSepPunct{\mcitedefaultmidpunct}
{\mcitedefaultendpunct}{\mcitedefaultseppunct}\relax
\EndOfBibitem
\bibitem[Tian \latin{et~al.}(2025)Tian, Zhu, Surman, Chen, and Sun]{tian2025achromatic}
Tian,~Z.; Zhu,~X.; Surman,~P.~A.; Chen,~Z.; Sun,~X.~W. An achromatic metasurface waveguide for augmented reality displays. \emph{Light: Science \& Applications} \textbf{2025}, \emph{14}, 94\relax
\mciteBstWouldAddEndPuncttrue
\mciteSetBstMidEndSepPunct{\mcitedefaultmidpunct}
{\mcitedefaultendpunct}{\mcitedefaultseppunct}\relax
\EndOfBibitem
\bibitem[Zang \latin{et~al.}(2024)Zang, Ren, Shi, Peng, Zheng, Zheng, Liu, Wang, Cheng, Liu, \latin{et~al.} others]{zang2024inverse}
Zang,~G.; Ren,~J.; Shi,~Y.; Peng,~D.; Zheng,~P.; Zheng,~K.; Liu,~Z.; Wang,~Z.; Cheng,~X.; Liu,~A.-Q.; others Inverse Design of Aberration-Corrected Hybrid Metalenses for Large Field of View Thermal Imaging Across the Entire Longwave Infrared Atmospheric Window. \emph{ACS nano} \textbf{2024}, \relax
\mciteBstWouldAddEndPunctfalse
\mciteSetBstMidEndSepPunct{\mcitedefaultmidpunct}
{}{\mcitedefaultseppunct}\relax
\EndOfBibitem
\bibitem[Fan \latin{et~al.}(2024)Fan, Cheng, Chen, Liu, Lu, Li, Jiang, Qin, and Dong]{fan2024integral}
Fan,~Z.-B.; Cheng,~Y.-F.; Chen,~Z.-M.; Liu,~X.; Lu,~W.-L.; Li,~S.-H.; Jiang,~S.-J.; Qin,~Z.; Dong,~J.-W. Integral imaging near-eye 3D display using a nanoimprint metalens array. \emph{eLight} \textbf{2024}, \emph{4}, 3\relax
\mciteBstWouldAddEndPuncttrue
\mciteSetBstMidEndSepPunct{\mcitedefaultmidpunct}
{\mcitedefaultendpunct}{\mcitedefaultseppunct}\relax
\EndOfBibitem
\bibitem[McClung \latin{et~al.}(2020)McClung, Samudrala, Torfeh, Mansouree, and Arbabi]{mcclung2020snapshot}
McClung,~A.; Samudrala,~S.; Torfeh,~M.; Mansouree,~M.; Arbabi,~A. Snapshot spectral imaging with parallel metasystems. \emph{Science advances} \textbf{2020}, \emph{6}, eabc7646\relax
\mciteBstWouldAddEndPuncttrue
\mciteSetBstMidEndSepPunct{\mcitedefaultmidpunct}
{\mcitedefaultendpunct}{\mcitedefaultseppunct}\relax
\EndOfBibitem
\bibitem[Kim \latin{et~al.}(2025)Kim, Lee, Yeo, Li, Kim, Kim, Badloe, Sun, Zhang, and Rho]{kim2025rapid}
Kim,~Y.; Lee,~J.; Yeo,~W.-H.; Li,~X.; Kim,~W.-S.; Kim,~Y.-K.; Badloe,~T.; Sun,~C.; Zhang,~H.~F.; Rho,~J. Rapid Polarization-Controlled Depth Sensing and Imaging with an Electrically Tunable Metalens. \emph{Nano Letters} \textbf{2025}, \relax
\mciteBstWouldAddEndPunctfalse
\mciteSetBstMidEndSepPunct{\mcitedefaultmidpunct}
{}{\mcitedefaultseppunct}\relax
\EndOfBibitem
\bibitem[Kuang \latin{et~al.}(2024)Kuang, Wang, Mo, Sun, Xia, and Yang]{kuang2024palm}
Kuang,~Y.; Wang,~S.; Mo,~B.; Sun,~S.; Xia,~K.; Yang,~Y. Palm vein imaging using a polarization-selective metalens with wide field-of-view and extended depth-of-field. \emph{npj Nanophotonics} \textbf{2024}, \emph{1}, 24\relax
\mciteBstWouldAddEndPuncttrue
\mciteSetBstMidEndSepPunct{\mcitedefaultmidpunct}
{\mcitedefaultendpunct}{\mcitedefaultseppunct}\relax
\EndOfBibitem
\bibitem[Shen \latin{et~al.}(2023)Shen, Zhao, Jin, Wang, Cao, and Yang]{shen2023monocular}
Shen,~Z.; Zhao,~F.; Jin,~C.; Wang,~S.; Cao,~L.; Yang,~Y. Monocular metasurface camera for passive single-shot 4D imaging. \emph{Nature Communications} \textbf{2023}, \emph{14}, 1035\relax
\mciteBstWouldAddEndPuncttrue
\mciteSetBstMidEndSepPunct{\mcitedefaultmidpunct}
{\mcitedefaultendpunct}{\mcitedefaultseppunct}\relax
\EndOfBibitem
\bibitem[Badloe \latin{et~al.}(2023)Badloe, Kim, Kim, Park, Barulin, Diep, Cho, Kim, Kim, Kim, \latin{et~al.} others]{badloe2023bright}
Badloe,~T.; Kim,~Y.; Kim,~J.; Park,~H.; Barulin,~A.; Diep,~Y.~N.; Cho,~H.; Kim,~W.-S.; Kim,~Y.-K.; Kim,~I.; others Bright-field and edge-enhanced imaging using an electrically tunable dual-mode metalens. \emph{ACS nano} \textbf{2023}, \emph{17}, 14678--14685\relax
\mciteBstWouldAddEndPuncttrue
\mciteSetBstMidEndSepPunct{\mcitedefaultmidpunct}
{\mcitedefaultendpunct}{\mcitedefaultseppunct}\relax
\EndOfBibitem
\bibitem[Fr{\"o}ch \latin{et~al.}(2025)Fr{\"o}ch, Colburn, Brady, Heide, Veeraraghavan, and Majumdar]{froch2025computational}
Fr{\"o}ch,~J.~E.; Colburn,~S.; Brady,~D.~J.; Heide,~F.; Veeraraghavan,~A.; Majumdar,~A. Computational imaging with meta-optics. \emph{Optica} \textbf{2025}, \emph{12}, 774--788\relax
\mciteBstWouldAddEndPuncttrue
\mciteSetBstMidEndSepPunct{\mcitedefaultmidpunct}
{\mcitedefaultendpunct}{\mcitedefaultseppunct}\relax
\EndOfBibitem
\bibitem[Aharoni \latin{et~al.}(2023)Aharoni, Federico, and Guo]{daniel_aharoni_2023_7844004}
Aharoni,~D.; Federico; Guo,~C. Aharoni-Lab/Miniscope-v4: Release for generating DOI. 2023; \url{https://doi.org/10.5281/zenodo.7844004}\relax
\mciteBstWouldAddEndPuncttrue
\mciteSetBstMidEndSepPunct{\mcitedefaultmidpunct}
{\mcitedefaultendpunct}{\mcitedefaultseppunct}\relax
\EndOfBibitem
\bibitem[Cao \latin{et~al.}(2023)Cao, Zhao, Li, Du, Zhang, Luo, Jiang, Davis, Zhou, de~la Zerda, \latin{et~al.} others]{cao2023optical}
Cao,~R.; Zhao,~J.; Li,~L.; Du,~L.; Zhang,~Y.; Luo,~Y.; Jiang,~L.; Davis,~S.; Zhou,~Q.; de~la Zerda,~A.; others Optical-resolution photoacoustic microscopy with a needle-shaped beam. \emph{Nature photonics} \textbf{2023}, \emph{17}, 89--95\relax
\mciteBstWouldAddEndPuncttrue
\mciteSetBstMidEndSepPunct{\mcitedefaultmidpunct}
{\mcitedefaultendpunct}{\mcitedefaultseppunct}\relax
\EndOfBibitem
\bibitem[Martins \latin{et~al.}(2020)Martins, Li, Li, Liang, Conteduca, Borges, Krauss, and Martins]{martins2020metalenses}
Martins,~A.; Li,~K.; Li,~J.; Liang,~H.; Conteduca,~D.; Borges,~B.-H.~V.; Krauss,~T.~F.; Martins,~E.~R. On metalenses with arbitrarily wide field of view. \emph{Acs Photonics} \textbf{2020}, \emph{7}, 2073--2079\relax
\mciteBstWouldAddEndPuncttrue
\mciteSetBstMidEndSepPunct{\mcitedefaultmidpunct}
{\mcitedefaultendpunct}{\mcitedefaultseppunct}\relax
\EndOfBibitem
\bibitem[Colburn and Majumdar(2020)Colburn, and Majumdar]{colburn2020metasurface}
Colburn,~S.; Majumdar,~A. Metasurface generation of paired accelerating and rotating optical beams for passive ranging and scene reconstruction. \emph{Acs Photonics} \textbf{2020}, \emph{7}, 1529--1536\relax
\mciteBstWouldAddEndPuncttrue
\mciteSetBstMidEndSepPunct{\mcitedefaultmidpunct}
{\mcitedefaultendpunct}{\mcitedefaultseppunct}\relax
\EndOfBibitem
\bibitem[Hao \latin{et~al.}(2024)Hao, Wang, Wang, Ding, Zhang, Pan, Rahman, Ling, Li, Tan, \latin{et~al.} others]{hao2024single}
Hao,~H.; Wang,~H.; Wang,~X.; Ding,~X.; Zhang,~S.; Pan,~C.-F.; Rahman,~M.~A.; Ling,~T.; Li,~H.; Tan,~J.; others Single-shot 3D imaging meta-microscope. \emph{Nano Letters} \textbf{2024}, \emph{24}, 13364--13373\relax
\mciteBstWouldAddEndPuncttrue
\mciteSetBstMidEndSepPunct{\mcitedefaultmidpunct}
{\mcitedefaultendpunct}{\mcitedefaultseppunct}\relax
\EndOfBibitem
\bibitem[Lee and Yun(2011)Lee, and Yun]{lee2011adaptive}
Lee,~W.; Yun,~S. Adaptive aberration correction of GRIN lenses for confocal endomicroscopy. \emph{Optics letters} \textbf{2011}, \emph{36}, 4608--4610\relax
\mciteBstWouldAddEndPuncttrue
\mciteSetBstMidEndSepPunct{\mcitedefaultmidpunct}
{\mcitedefaultendpunct}{\mcitedefaultseppunct}\relax
\EndOfBibitem
\bibitem[Easum \latin{et~al.}(2016)Easum, Campbell, Nagar, and Werner]{easum2016analytical}
Easum,~J.~A.; Campbell,~S.~D.; Nagar,~J.; Werner,~D.~H. Analytical surrogate model for the aberrations of an arbitrary GRIN lens. \emph{Optics express} \textbf{2016}, \emph{24}, 17805--17818\relax
\mciteBstWouldAddEndPuncttrue
\mciteSetBstMidEndSepPunct{\mcitedefaultmidpunct}
{\mcitedefaultendpunct}{\mcitedefaultseppunct}\relax
\EndOfBibitem
\bibitem[Khorasaninejad \latin{et~al.}(2016)Khorasaninejad, Chen, Devlin, Oh, Zhu, and Capasso]{khorasaninejad2016metalenses}
Khorasaninejad,~M.; Chen,~W.~T.; Devlin,~R.~C.; Oh,~J.; Zhu,~A.~Y.; Capasso,~F. Metalenses at visible wavelengths: Diffraction-limited focusing and subwavelength resolution imaging. \emph{Science} \textbf{2016}, \emph{352}, 1190--1194\relax
\mciteBstWouldAddEndPuncttrue
\mciteSetBstMidEndSepPunct{\mcitedefaultmidpunct}
{\mcitedefaultendpunct}{\mcitedefaultseppunct}\relax
\EndOfBibitem
\bibitem[Pu \latin{et~al.}(2017)Pu, Li, Guo, Ma, and Luo]{pu2017nanoapertures}
Pu,~M.; Li,~X.; Guo,~Y.; Ma,~X.; Luo,~X. Nanoapertures with ordered rotations: symmetry transformation and wide-angle flat lensing. \emph{Optics Express} \textbf{2017}, \emph{25}, 31471--31477\relax
\mciteBstWouldAddEndPuncttrue
\mciteSetBstMidEndSepPunct{\mcitedefaultmidpunct}
{\mcitedefaultendpunct}{\mcitedefaultseppunct}\relax
\EndOfBibitem
\bibitem[Liang \latin{et~al.}(2019)Liang, Martins, Borges, Zhou, Martins, Li, and Krauss]{liang2019high}
Liang,~H.; Martins,~A.; Borges,~B.-H.~V.; Zhou,~J.; Martins,~E.~R.; Li,~J.; Krauss,~T.~F. High performance metalenses: numerical aperture, aberrations, chromaticity, and trade-offs. \emph{Optica} \textbf{2019}, \emph{6}, 1461--1470\relax
\mciteBstWouldAddEndPuncttrue
\mciteSetBstMidEndSepPunct{\mcitedefaultmidpunct}
{\mcitedefaultendpunct}{\mcitedefaultseppunct}\relax
\EndOfBibitem
\bibitem[Bayati \latin{et~al.}(2020)Bayati, Pestourie, Colburn, Lin, Johnson, and Majumdar]{bayati2020inverse}
Bayati,~E.; Pestourie,~R.; Colburn,~S.; Lin,~Z.; Johnson,~S.~G.; Majumdar,~A. Inverse designed metalenses with extended depth of focus. \emph{ACS photonics} \textbf{2020}, \emph{7}, 873--878\relax
\mciteBstWouldAddEndPuncttrue
\mciteSetBstMidEndSepPunct{\mcitedefaultmidpunct}
{\mcitedefaultendpunct}{\mcitedefaultseppunct}\relax
\EndOfBibitem
\bibitem[Fr{\"o}ch \latin{et~al.}(2025)Fr{\"o}ch, Chakravarthula, Sun, Tseng, Colburn, Zhan, Miller, Wirth-Singh, Tanguy, Han, \latin{et~al.} others]{froch2025beating}
Fr{\"o}ch,~J.~E.; Chakravarthula,~P.; Sun,~J.; Tseng,~E.; Colburn,~S.; Zhan,~A.; Miller,~F.; Wirth-Singh,~A.; Tanguy,~Q.~A.; Han,~Z.; others Beating spectral bandwidth limits for large aperture broadband nano-optics. \emph{Nature communications} \textbf{2025}, \emph{16}, 3025\relax
\mciteBstWouldAddEndPuncttrue
\mciteSetBstMidEndSepPunct{\mcitedefaultmidpunct}
{\mcitedefaultendpunct}{\mcitedefaultseppunct}\relax
\EndOfBibitem
\bibitem[Yang \latin{et~al.}(2023)Yang, Wei, Zhao, Li, Zhang, Li, Li, Jing, Li, Wang, \latin{et~al.} others]{yang2023realizing}
Yang,~S.; Wei,~Q.; Zhao,~R.; Li,~X.; Zhang,~X.; Li,~Y.; Li,~J.; Jing,~X.; Li,~X.; Wang,~Y.; others Realizing depth measurement and edge detection based on a single metasurface. \emph{Nanophotonics} \textbf{2023}, \emph{12}, 3385--3393\relax
\mciteBstWouldAddEndPuncttrue
\mciteSetBstMidEndSepPunct{\mcitedefaultmidpunct}
{\mcitedefaultendpunct}{\mcitedefaultseppunct}\relax
\EndOfBibitem
\bibitem[Jin \latin{et~al.}(2019)Jin, Zhang, and Guo]{jin2019metasurface}
Jin,~C.; Zhang,~J.; Guo,~C. Metasurface integrated with double-helix point spread function and metalens for three-dimensional imaging. \emph{Nanophotonics} \textbf{2019}, \emph{8}, 451--458\relax
\mciteBstWouldAddEndPuncttrue
\mciteSetBstMidEndSepPunct{\mcitedefaultmidpunct}
{\mcitedefaultendpunct}{\mcitedefaultseppunct}\relax
\EndOfBibitem
\bibitem[Fr{\"o}ch \latin{et~al.}(2022)Fr{\"o}ch, Colburn, Zhan, Han, Fang, Saxena, Huang, B{\"o}hringer, and Majumdar]{fröch2022dual}
Fr{\"o}ch,~J.~E.; Colburn,~S.; Zhan,~A.; Han,~Z.; Fang,~Z.; Saxena,~A.; Huang,~L.; B{\"o}hringer,~K.~F.; Majumdar,~A. Dual band computational infrared spectroscopy via large aperture meta-optics. \emph{ACS Photonics} \textbf{2022}, \emph{10}, 986--992\relax
\mciteBstWouldAddEndPuncttrue
\mciteSetBstMidEndSepPunct{\mcitedefaultmidpunct}
{\mcitedefaultendpunct}{\mcitedefaultseppunct}\relax
\EndOfBibitem
\bibitem[Barulin \latin{et~al.}(2025)Barulin, Barulina, Oh, Jo, Park, Park, Kye, Kim, Yoo, Kim, \latin{et~al.} others]{barulin2025axially}
Barulin,~A.; Barulina,~E.; Oh,~D.~K.; Jo,~Y.; Park,~H.; Park,~S.; Kye,~H.; Kim,~J.; Yoo,~J.; Kim,~J.; others Axially multifocal metalens for 3D volumetric photoacoustic imaging of neuromelanin in live brain organoid. \emph{Science Advances} \textbf{2025}, \emph{11}, eadr0654\relax
\mciteBstWouldAddEndPuncttrue
\mciteSetBstMidEndSepPunct{\mcitedefaultmidpunct}
{\mcitedefaultendpunct}{\mcitedefaultseppunct}\relax
\EndOfBibitem
\bibitem[Zhao \latin{et~al.}(2022)Zhao, Winetraub, Du, Van~Vleck, Ichimura, Huang, Aasi, Sarin, and de~la Zerda]{zhao2022flexible}
Zhao,~J.; Winetraub,~Y.; Du,~L.; Van~Vleck,~A.; Ichimura,~K.; Huang,~C.; Aasi,~S.~Z.; Sarin,~K.~Y.; de~la Zerda,~A. Flexible method for generating needle-shaped beams and its application in optical coherence tomography. \emph{Optica} \textbf{2022}, \emph{9}, 859--867\relax
\mciteBstWouldAddEndPuncttrue
\mciteSetBstMidEndSepPunct{\mcitedefaultmidpunct}
{\mcitedefaultendpunct}{\mcitedefaultseppunct}\relax
\EndOfBibitem
\bibitem[Audhkhasi \latin{et~al.}(2025)Audhkhasi, Xie, Froch, and Majumdar]{audhkhasi2025single}
Audhkhasi,~R.; Xie,~N.; Froch,~J.~E.; Majumdar,~A. Single-shot Multispectral Imaging via a Chromatic Metalens Array. \emph{ACS Photonics} \textbf{2025}, \emph{12}, 2761--2766\relax
\mciteBstWouldAddEndPuncttrue
\mciteSetBstMidEndSepPunct{\mcitedefaultmidpunct}
{\mcitedefaultendpunct}{\mcitedefaultseppunct}\relax
\EndOfBibitem
\end{mcitethebibliography}
\end{document}